\documentclass[final,3p,times]{elsarticle}

\usepackage{multirow,setspace,times,amssymb,amsmath,graphicx,color,rotating,subfigure,url}
\usepackage{lineno}
\usepackage{natbib}

\journal{Physica A}

\begin{document}

\begin{frontmatter}

\title{Horizontal visibility graphs transformed from fractional Brownian motions:\\Topological properties versus Hurst index}
\author[BS,SS]{Wen-Jie Xie}
\author[BS,SS,RCE]{Wei-Xing Zhou \corref{cor}}
\cortext[cor]{Corresponding author.}
\ead{wxzhou@ecust.edu.cn} %
\ead[url]{http://rce.ecust.edu.cn/index.php/en/wxzhou}%

\address[BS]{School of Business, East China University of Science and Technology, Shanghai 200237, China}
\address[SS]{School of Science, East China University of Science and Technology, Shanghai 200237, China}
\address[RCE]{Research Center for Econophysics, East China University of Science and Technology, Shanghai 200237, China}

\begin{abstract}
Nonlinear time series analysis aims at understanding the dynamics of stochastic or chaotic processes. In recent years, quite a few methods have been proposed to transform a single time series to a complex network so that the dynamics of the process can be understood by investigating the topological properties of the network. We study the topological properties of horizontal visibility graphs constructed from fractional Brownian motions with different Hurst index $H\in(0,1)$. Special attention has been paid to the impact of Hurst index on the topological properties. It is found that the clustering coefficient $C$ decreases when $H$ increases. We also found that the mean length $L$ of the shortest paths increases exponentially with $H$ for fixed length $N$ of the original time series. In addition, $L$ increases linearly with respect to $N$ when $H$ is close to 1 and in a logarithmic form when $H$ is close to 0. Although the occurrence of different motifs changes with $H$, the motif rank pattern remains unchanged for different $H$. Adopting the node-covering box-counting method, the horizontal visibility graphs are found to be fractals and the fractal dimension $d_B$ decreases with $H$. Furthermore, the Pearson coefficients of the networks are positive and the degree-degree correlations increase with the degree, which indicate that the horizontal visibility graphs are assortative. With the increase of $H$, the Pearson coefficient decreases first and then increases, in which the turning point is around $H=0.6$. The presence of both fractality and assortativity in the horizontal visibility graphs converted from fractional Brownian motions is different from many cases where fractal networks are usually disassortative.
\end{abstract}


\end{frontmatter}

\section{Introduction}
\label{S1:Introduction}

Nonlinear time series analysis aims at understanding the dynamics of the underlying process. In recent years, numerous methods have been proposed to transform a single time series to a complex network \cite{Small-Zhang-Xu-2009-LNICST,Donner-Small-Donges-Marwan-Zou-Xiang-Kurths-2011-IJBC}. In this way, the dynamics of the process can be partially understood through the investigation of the network's topological properties. This direction is promising because there are a large number of tools developed in the field of network sciences \cite{Albert-Barabasi-2002-RMP,Newman-2003-SIAMR,Boccaletti-Latora-Moreno-Chavez-Hwang-2006-PR}. Different transforming algorithms result in different networks, such as cycle networks based on the local extrema and their distances in the phase space \cite{Zhang-Small-2006-PRL,Zhang-Sun-Luo-Zhang-Nakamura-Small-2008-PD,Zhang-Zhou-Wang-2010-PProc}, $n$-tuple networks \cite{Li-Wang-2006-CSB,Li-Wang-2007-PA}, space state networks based on conformational fluctuations \cite{Li-Yang-Komatsuzak-2008-PNAS}, visibility graphs based on the visibility of nodes \cite{Lacasa-Luque-Ballesteros-Luque-Nuno-2008-PNAS,Luque-Lacasa-Ballesteros-Luque-2009-PRE}, nearest neighbor networks in the phase space \cite{Xu-Zhang-Small-2008-PNAS,Gao-Jin-2009-Chaos,Liu-Zhou-2010-JPA}, segment correlation networks \cite{Yang-Yang-2008-PA,Gao-Jin-2009-PRE}, temporal graphs \cite{Shirazi-Jafari-Davoudi-Peinke-Tabar-Sahimi-2009-JSM,Kostakos-2009-PA}, and recurrence networks \cite{Marwan-Donges-Zou-Donner-Kurths-2009-PLA,Donner-Zou-Donges-Marwan-Kurths-2010-PRE,Donner-Zou-Donges-Marwan-Kurths-2010-NJP}.
In a recent work, a method is proposed to convert an ensemble of sequences of symbols into a weighted directed network whose nodes are motifs, while the directed links and their weights are defined from statistically significant co-occurrences of two motifs in the same sequence \cite{Sinatra-Condorelli-Latora-2010-PRL}. This method can also be applied to time series analysis.

Among the aforementioned methods, the visibility algorithm has attracted most applications from diverse fields, including stock market indices \cite{Ni-Jiang-Zhou-2009-PLA,Qian-Jiang-Zhou-2010-JPA}, human stride intervals \cite{Lacasa-Luque-Luque-Nuno-2009-EPL}, occurrence of hurricanes in the United States \cite{Elsner-Jagger-Fogarty-2009-GRL}, foreign exchange rates \cite{Yang-Wang-Yang-Mang-2009-PA}, energy dissipation rates in three-dimensional fully developed turbulence \cite{Liu-Zhou-Yuan-2010-PA}, human heartbeat dynamics \cite{Shao-2010-APL,Dong-Li-2010-APL}, electroencephalogram series \cite{Ahmadlou-Adeli-Adeli-2010-JNT}, binary sequences \cite{Ahadpour-Sadra-2010-XXX}, interevent time of book loans \cite{Fan-Guo-Zha-2010-XXX}, and daily streamflow series \cite{Tang-Liu-Liu-2010-MPLB}. The visibility algorithm transforms a time series $\{x_{i}\}_{i = 1,...,N}$ into a visibility graph $G = \langle V,E\rangle$, where $V = \{v_{i}\}_{i = 1,...,N}$ is the set of vertex with the vertex $v_{i}$ corresponding to the data point $x_{i}$ and $E = \{e_{i,j}\}_{i,j = 1,...,N}$ is the adjacent matrix of the visibility graph whose element $e_{i,j} = 1$ if the following geometrical criterion is fulfilled:
\begin{equation}
   \frac{x_{i}-x_{n}}{i-n}>\frac{x_{i}-x_{j}}{i-j},~\forall{n}|i<n<j,
   \label{Eq:VG}
\end{equation}
and $e_{ij}=0$ otherwise. The degree distribution of the visibility graph contains information of the original time series. For random series extracted from a uniform distribution in $[0, 1]$, the degree distribution has an exponential tail $P(k)\sim e^{k/k_{0}}$ \cite{Lacasa-Luque-Ballesteros-Luque-Nuno-2008-PNAS}. In contrast, for fractional Brownian motions (FBMs), the degree distributions have power-law tails $P(k)\sim k^{-\alpha}$, whose exponent $\alpha$ decreases linearly with the Hurst index $H$ \cite{Lacasa-Luque-Luque-Nuno-2009-EPL,Ni-Jiang-Zhou-2009-PLA}.

Recently, a variant of the visibility algorithm has been proposed, which transforms time series into horizontal visibility graphs (HVGs) \cite{Luque-Lacasa-Ballesteros-Luque-2009-PRE}. We note that the horizontal visibility algorithm is a variant of the visibility algorithm. A graph is an HVG if and only if it is outerplanar and has a Hamilton path \cite{Gutin-Mansour-Severini-2010-XXX}. It has been proven that the degree distribution of any HVG mapped from random series without temporal correlations has an exponential form $P(k)=(3/4)e^{-k\ln(3/2)}$, which allows us to distinguish chaotic series from independent and identically distributed (i.i.d.) time series \cite{Luque-Lacasa-Ballesteros-Luque-2009-PRE}. Furthermore, numerical simulations show that an HVG mapped from chaotic or correlated time series has exponential degree distribution
\begin{equation}
 P(k) \sim e^{-\lambda{k}},
 \label{Eq:HVG:PDF:k}
\end{equation}
in which a chaotic process has $\lambda<\ln(3/2)$ and a correlated stochastic series has $\lambda>\ln(3/2)$, separated by the i.i.d. case with $\lambda=\ln(3/2)$ \cite{Lacasa-Toral-2010-PRE}.

In this work, we investigate the topological properties of HVGs mapped from fractional Brownian motions with different Hurst indexes. Special attention is paid to the influence of the Hurst index of the fractional Brownian motion on the topological properties of the associated HVG. Specifically, the degree distribution, the clustering coefficient, the mean length of the shortest paths, the motif distribution, the fractal nature, and the mixing behavior of the HVGs are numerically studies. The most striking result is that the HVGs possesses both fractal and assortative features, which is different from the usual conclusion that fractal networks are disassortative \cite{Yook-Radicchi-MeyerOrtmanns-2005-PRE,Zhang-Zhou-Zou-2007-EPJB}.

The paper is organized as follows. Section \ref{S1:BasicProperties} studies the basic properties of the horizontal visibility graphs mapped from FBMs. Section \ref{S1:Fractal} investigates the fractal nature of the HVGs using the node-covering box-counting method \cite{Song-Havlin-Makse-2005-Nature}. Section \ref{S1:Assortative} explores the mixing pattern of the HVGs. Section \ref{S1:Conclusion} summarizes our findings.

\section{Basic topological properties of the horizontal visibility graphs}
\label{S1:BasicProperties}

\subsection{Illustrative examples of the horizontal visibility algorithm and HVGs}
\label{S2:IllustrativeExample}

\begin{figure}[htb]
  \centering
  \includegraphics[width=6cm]{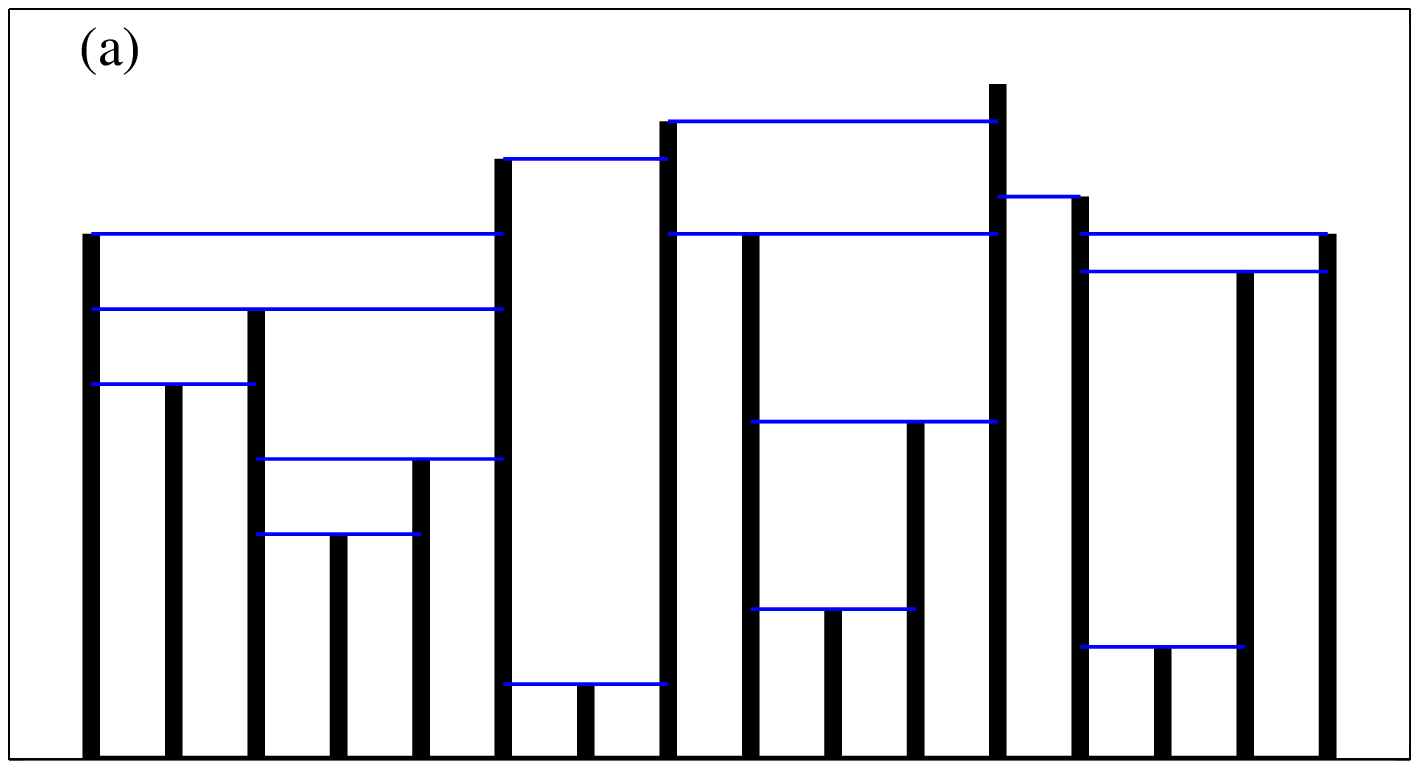}
  \includegraphics[width=6cm]{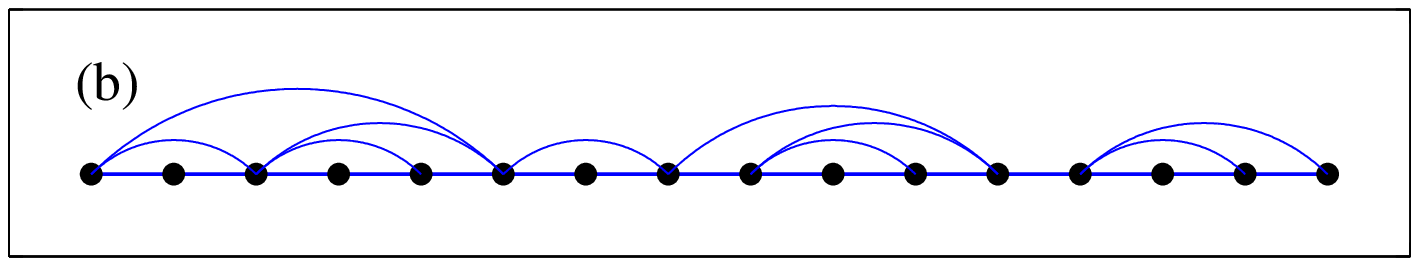}
  \caption{\label{Fig:HVG:Algo} (Color online) The horizontal visibility algorithm. (a) A time series with horizontal visibility links. (b) The horizontal visibility graph mapped from the time series in (a).}
\end{figure}

The horizontal visibility algorithm transforms a time series $\{x_{i}\}_{i = 1,...,N}$ into a horizontal visibility graph $G = \langle V, E\rangle$, where $V = \{v_{i}\}$ is a set of nodes corresponding to data points $\{x_{i}\}$ and $E = \{e_{i,j}\}$ is the adjacent matrix of graph $G$, whose element $e_{i,j} = 1$ if every points $x_{n}$ between $x_{i}$ and $x_{j}$ fulfill the following condition \cite{Luque-Lacasa-Ballesteros-Luque-2009-PRE}:
\begin{equation}
   x_{i}, x_{j} >  x_{n},~\forall{n}|i<n<j.
    \label{Eq:HVG}
\end{equation}
Otherwise, $e_{i,j} = 0$, stating that point $x_{i}$ is not horizontally visible to point $x_{j}$ and the corresponding vertices $v_{i}$ and $v_{j}$ are not connected. An illustrative example is shown in Fig.~\ref{Fig:HVG:Algo}. Figure \ref{Fig:HVG:Algo}(a) plots a simple time series containing 16 data points by vertical thick bars, where the height of each bar is equal to the value of the corresponding data point. According to the horizontal visibility algorithm, we link every two bars when all the bars between them are lower than both of them and a horizontal link is drawn between these two bars that are visible to each other in the horizontal direction. The resultant horizontal visibility graph is shown in Fig.~\ref{Fig:HVG:Algo}(b).

\begin{figure}[htb]
  \centering
  \includegraphics[width=5.3cm]{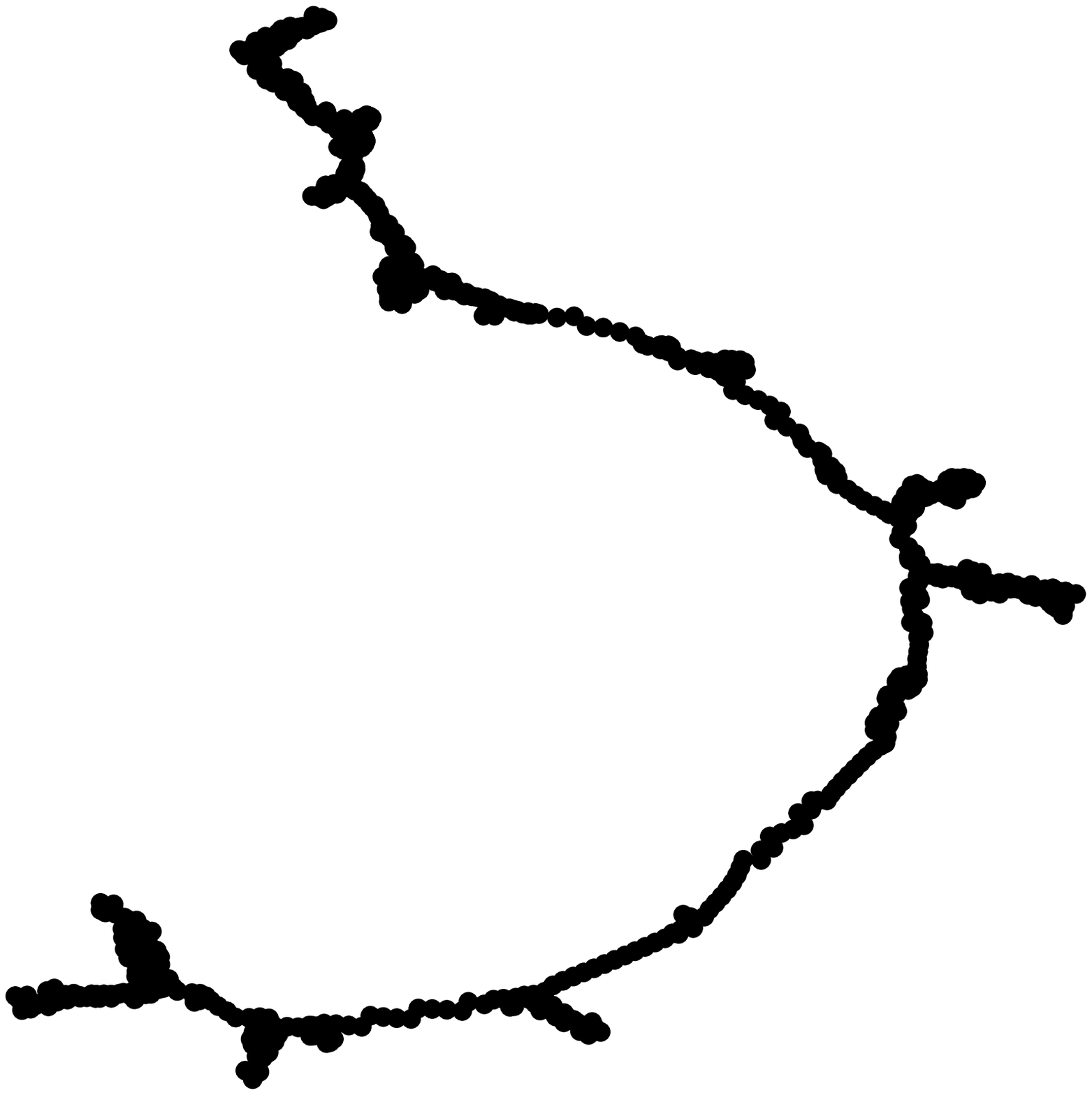}
  \includegraphics[width=5.3cm]{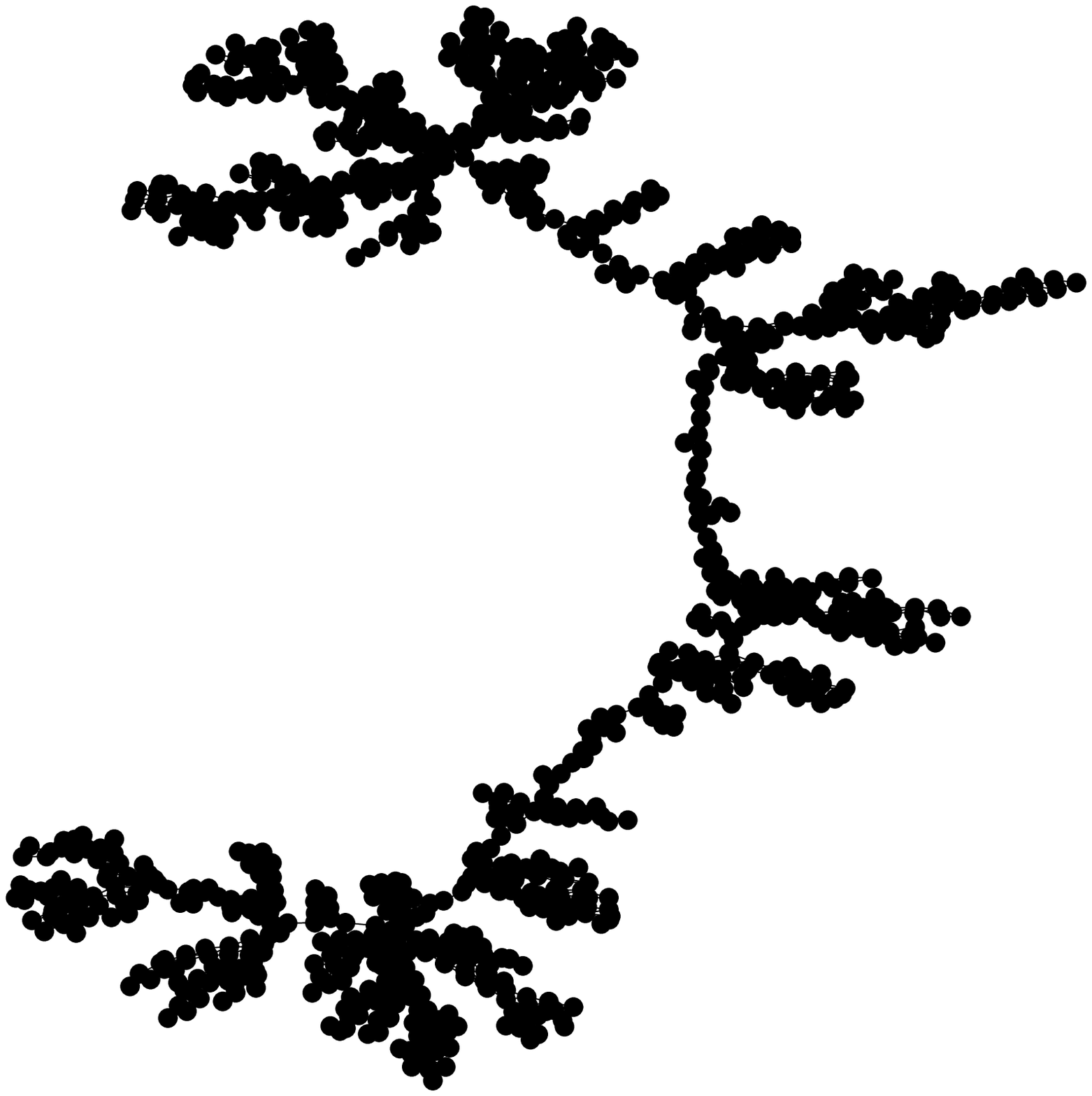}
  \includegraphics[width=5.3cm]{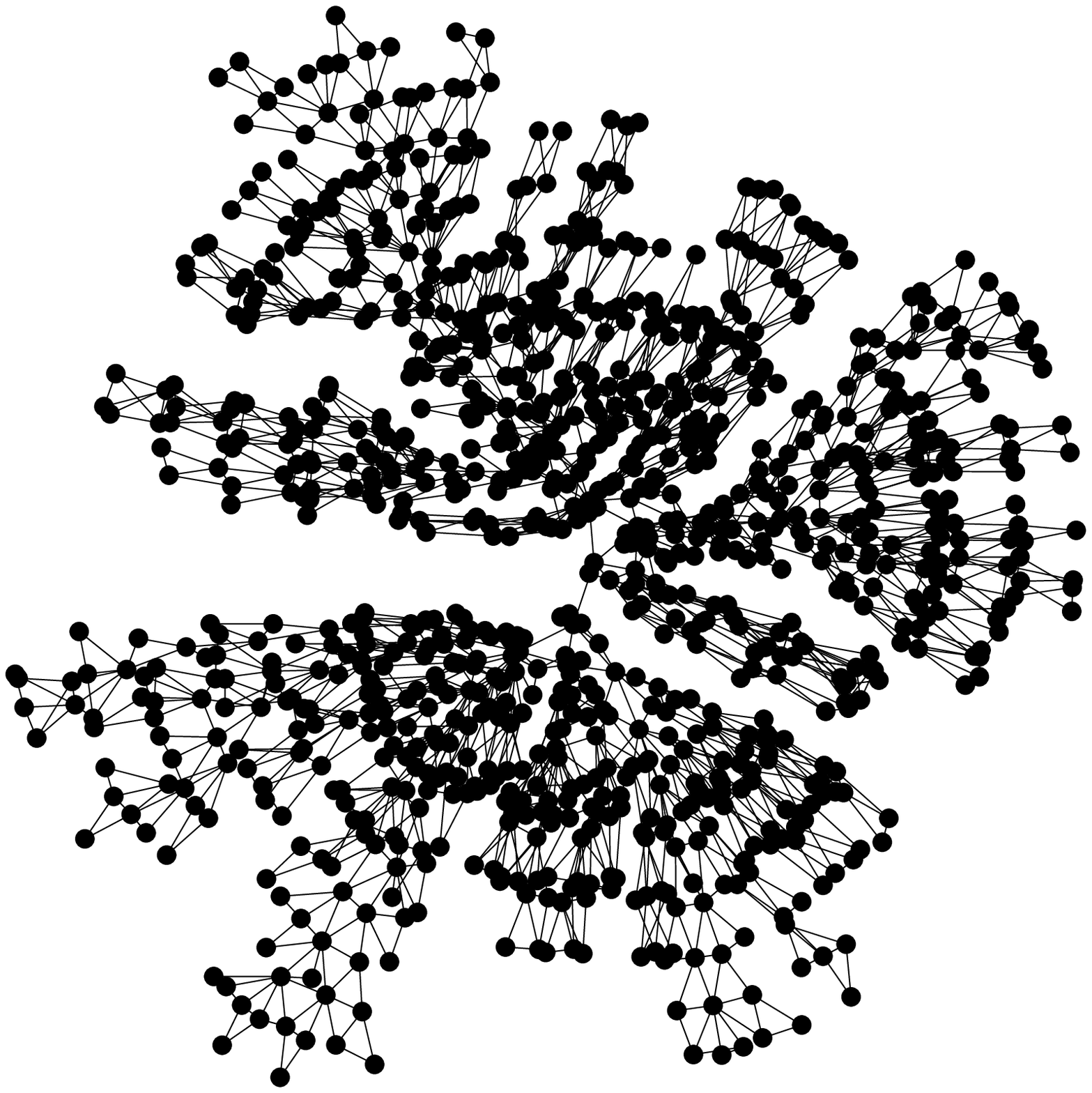}
  \includegraphics[width=5.3cm]{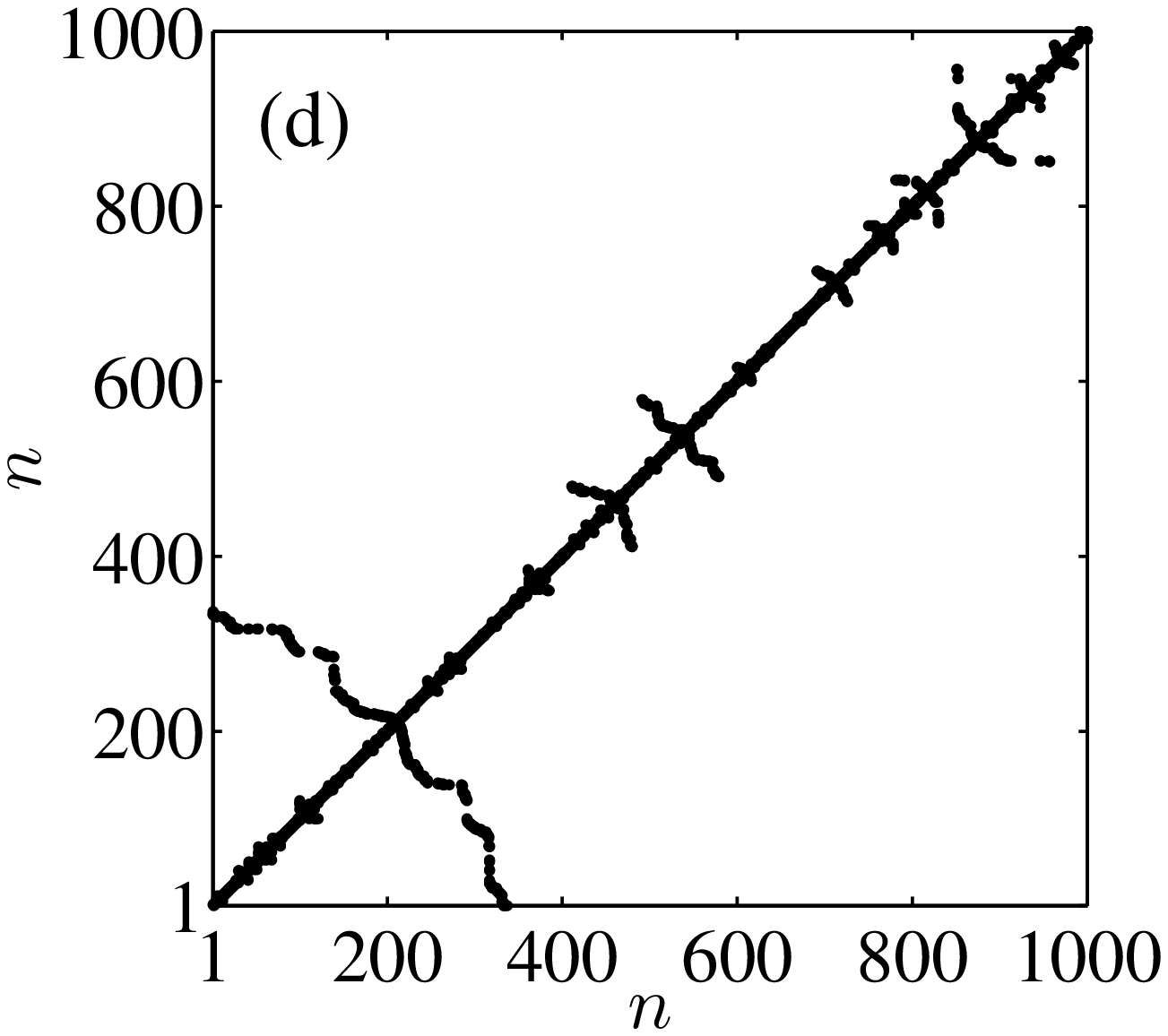}
  \includegraphics[width=5.3cm]{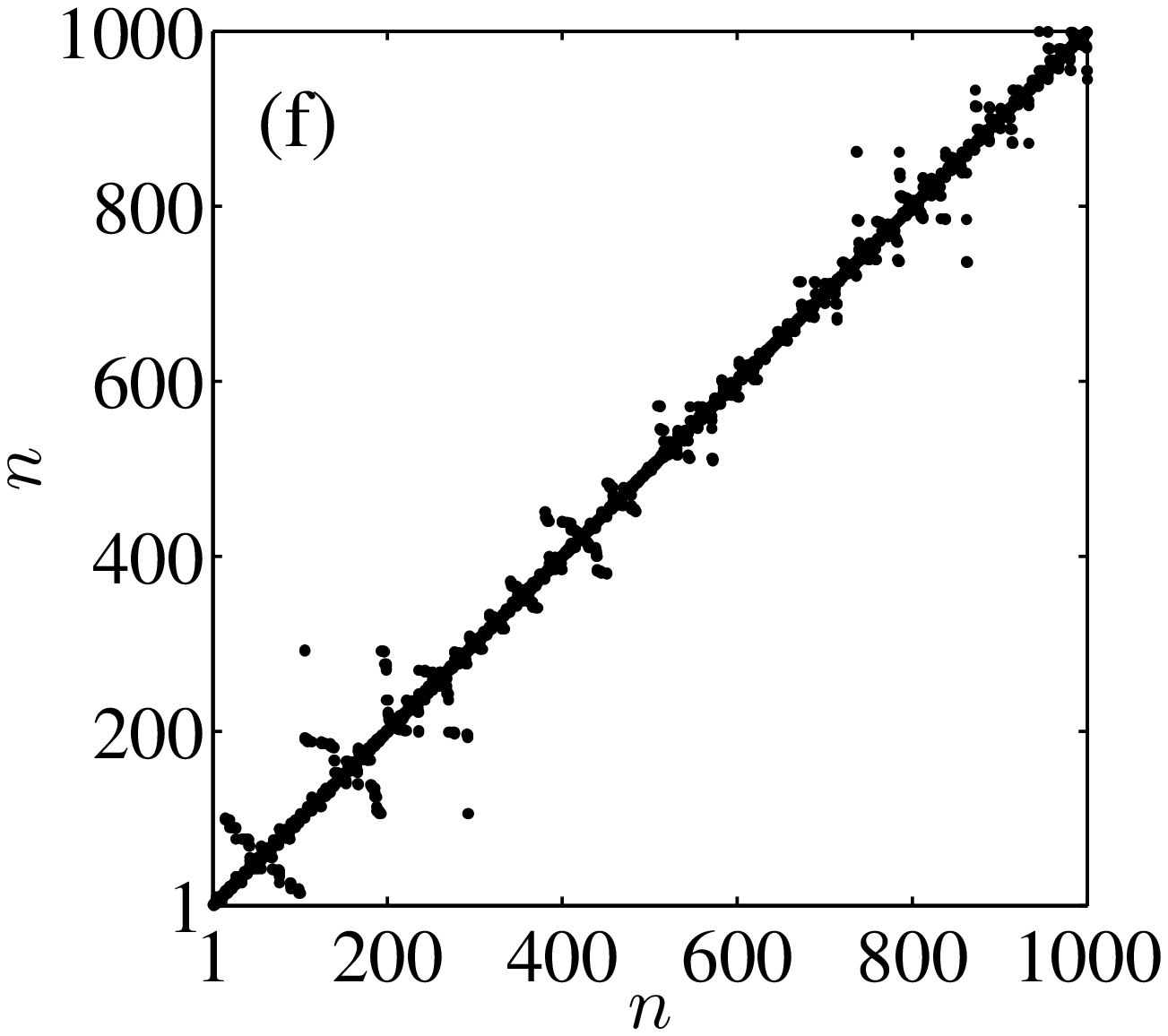}
  \includegraphics[width=5.3cm]{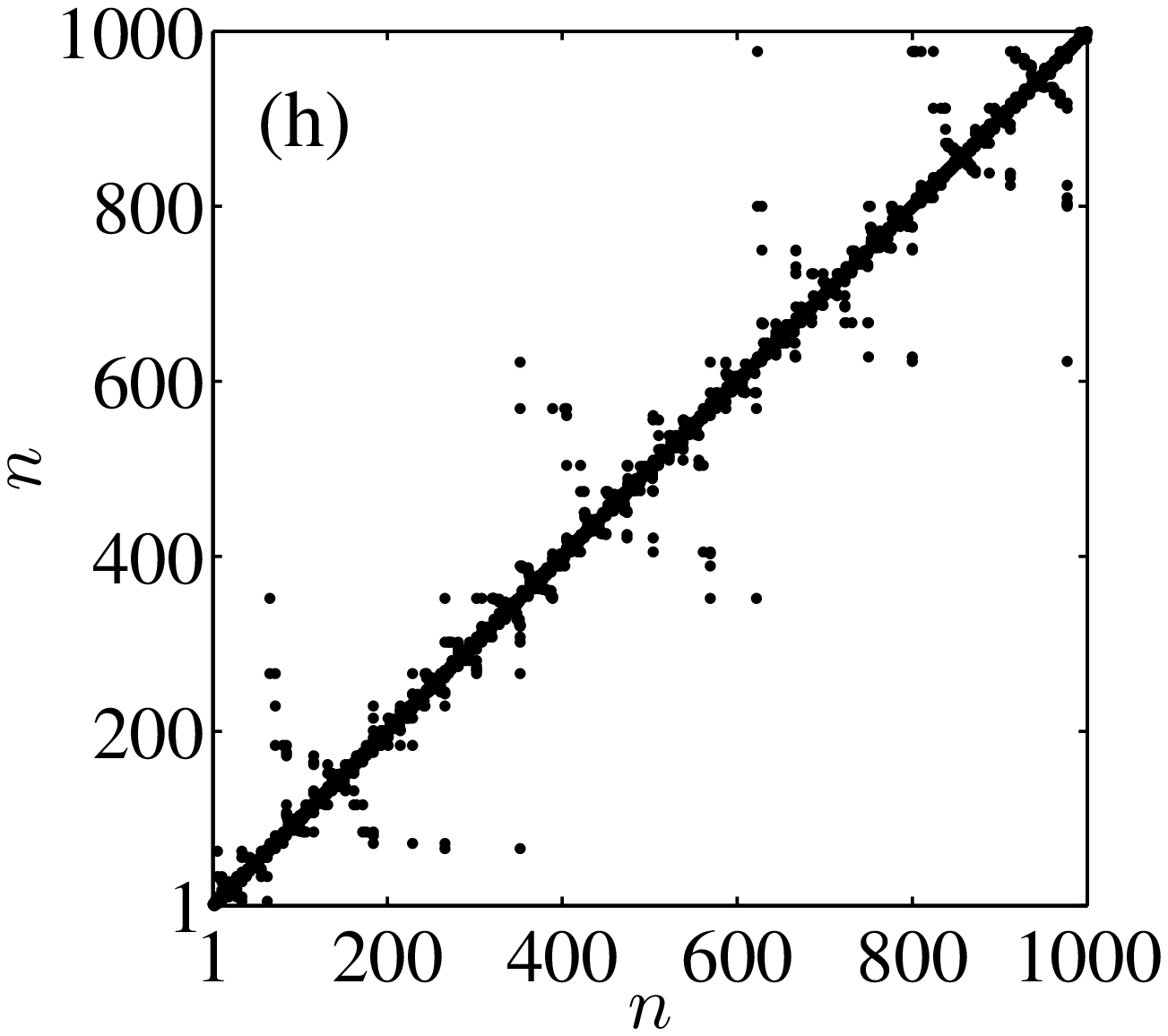}
  \vskip -9.6cm \hskip -15cm  {(a)}
  \vskip -0.4cm \hskip -5.0cm {(b)}
  \vskip -0.4cm \hskip  6.3cm {(c)}
\vskip 9cm
  \caption{\label{Fig:HVG:Examples} Illustrative examples of HVGs. (a-c) HVGs mapped from FBM series of length $N=1000$ with Hurst indexes $H = 0.9$, 0.5 and 0.1. (d-f) Adjacent matrices of the HVGs in (a-c).}
\end{figure}

Figure \ref{Fig:HVG:Examples} illustrates three typical HVGs mapped from FBMs with different Hurst indexes $H = 0.9$, 0.5 and 0.1 and the corresponding adjacent matrices. With the decrease of $H$, the HVG contains more loops and exhibits more fine structures and the diameter becomes shorter. We can conjecture from Fig.~\ref{Fig:HVG:Examples}(a) that the mean length of the shortest paths is proportional to the size $N$ of the HVG for large $H$. By contrast, the HVGs for small $H$ may exhibit small-world nature. In addition, the HVG mapped from an FBM with smaller Hurst index is more homogenous. These HVGs are reminiscent of fractal patterns such as viscous fingers, cracks, and diffusion-limited aggregates, implying that these graphs might have fractal nature. It is evident that the topological properties of the HVGs are strongly influenced by the Hurst indexes of the FBMs, which will be investigated in detail through numerical simulations. Consistently, the adjacent matrix varies in respect to the Hurst index.

\subsection{Degree distribution}
\label{S2:DegreeDistribution}

\begin{figure}[htb]
\centering
\includegraphics[width=8cm]{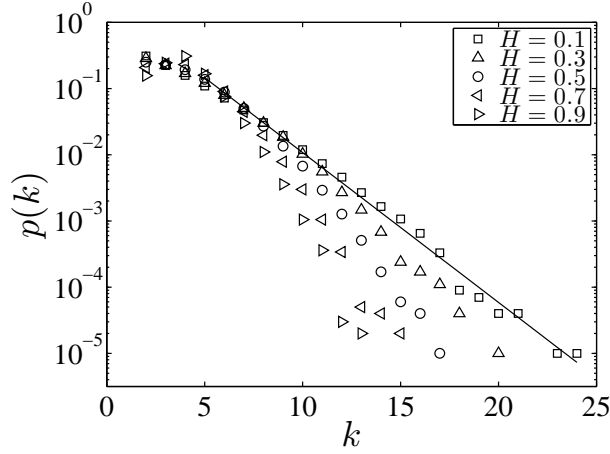}
 \caption{\label{Fig:HVG:Pk} Empirical degree distributions of the horizontal visibility graphs converted from FBM series with different Hurst indexes $H = 0.1$, 0.3, 0.5, 0.7, and 0.9. Each FBM series contains $10^5$ data points. The distributions have exponential tails and the parameter $\lambda$ increases with $H$. The solid line is the linear fit for $H = 0.1$.}
\end{figure}

Degree distribution $P(k)$ is one of the most important characteristic properties of complex networks. The degree distributions of the HVGs of correlated and uncorrelated stochastic and chaotic time series are exponential \cite{Luque-Lacasa-Ballesteros-Luque-2009-PRE,Lacasa-Toral-2010-PRE}. In addition, the visibility graphs of uncorrelated time series and fractional Brownian motions have exponential degree distributions and power-law tails, respectively \cite{Lacasa-Luque-Ballesteros-Luque-Nuno-2008-PNAS,Lacasa-Luque-Luque-Nuno-2009-EPL,Ni-Jiang-Zhou-2009-PLA}. A natural conjecture is that the degree distributions of the HVGs transformed from fractional Brownian motions might have power-law tails. We study the problem numerically.  Figure \ref{Fig:HVG:Pk} shows the degree distributions for the Hurst indexes $H = 0.1$, 0.3, 0.5, 0.7 and 0.9. It is found that the degree distributions have exponential tails as in Eq.~(\ref{Eq:HVG:PDF:k}) and the parameter $\lambda$ increases with $H$.

\subsection{Clustering coefficient}
\label{S2:ClusteringCoefficient}

\begin{figure}[htb]
  \centering
  \includegraphics[width=8cm]{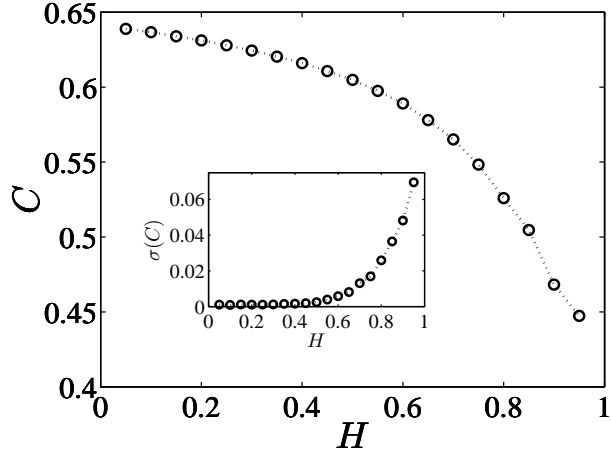}
  \caption{\label{Fig:HVG:C:H} Dependence of the clustering coefficient $C$ of the HVGs against the Hurst index $H$ of the original FBMs. Each FBM time series contains $5\times10^3$ data points. For each $H$, the simulation is repeated for 100 times and the averages of $C$ are shown. The inset shows the variance of $C$.}
\end{figure}

We now study the dependence of the clustering coefficient $C$ of the HVG on the Hurst index of the FBM. The clustering coefficient $C$ of a network defined as follows:
\begin{equation}
  C = \frac{\sum_{v}C_{v}}{N},
\end{equation}
where
\begin{equation}
  C_{v} = \frac{\sum_{i,j}^{N}e_{v,i}e_{i,j}e_{j,v}}{k_{v}(k_{v}-1)},
\end{equation}
where $k_{v}$ is degree of node $v$. The average of local clustering coefficient $C$ states the probability of two neighboring nodes of any node are neighbors too. In other words, the clustering coefficient describes the occurrence frequency of triangles in the network \cite{Newman-2003-SIAMR}. Figure \ref{Fig:HVG:C:H} shows that $C$ decreases with $H$ and the decrease speed also increases with $H$. This observation is evident according to Fig.~\ref{Fig:HVG:Examples}, which shows that there are more triangles in the HVG when $H$ decreases from 0.9 to 0.1. For small Hurst indexes, the clustering coefficients are large.

\subsection{Mean length of the shortest paths}
\label{S2:MSP}

The mean length of the shortest paths of a network can be calculated as follows:
\begin{equation}
  L(N) = \frac{2}{N(N-1)}\sum_{i=1}^{N-1}\sum_{j=i+1}^N d(i,j),
\end{equation}
where $d(i,j)$ is the length of the shortest path between node $i$ and $j$. For the HVG mapped from i.i.d. time series, the mean length of the shortest paths can be approximately derived:
\begin{equation}
 L(N)=2\ln{N} + 2(\gamma-1)+O(1/N),
 \label{Eq:HVG:L:N:iid}
\end{equation}
where $\gamma$ is the Euler-Mascheroni constant \cite{Luque-Lacasa-Ballesteros-Luque-2009-PRE}. However, it is a hard task to obtain an analytic expression of $L$ for correlated series \cite{Lacasa-Toral-2010-PRE}. Similarly, it is also difficult to derive the expression of $L$ for the HVGs converted from FBMs. We thus study this problem by numerical simulations.

\begin{figure}[htb]
  \centering
  \includegraphics[width=8cm]{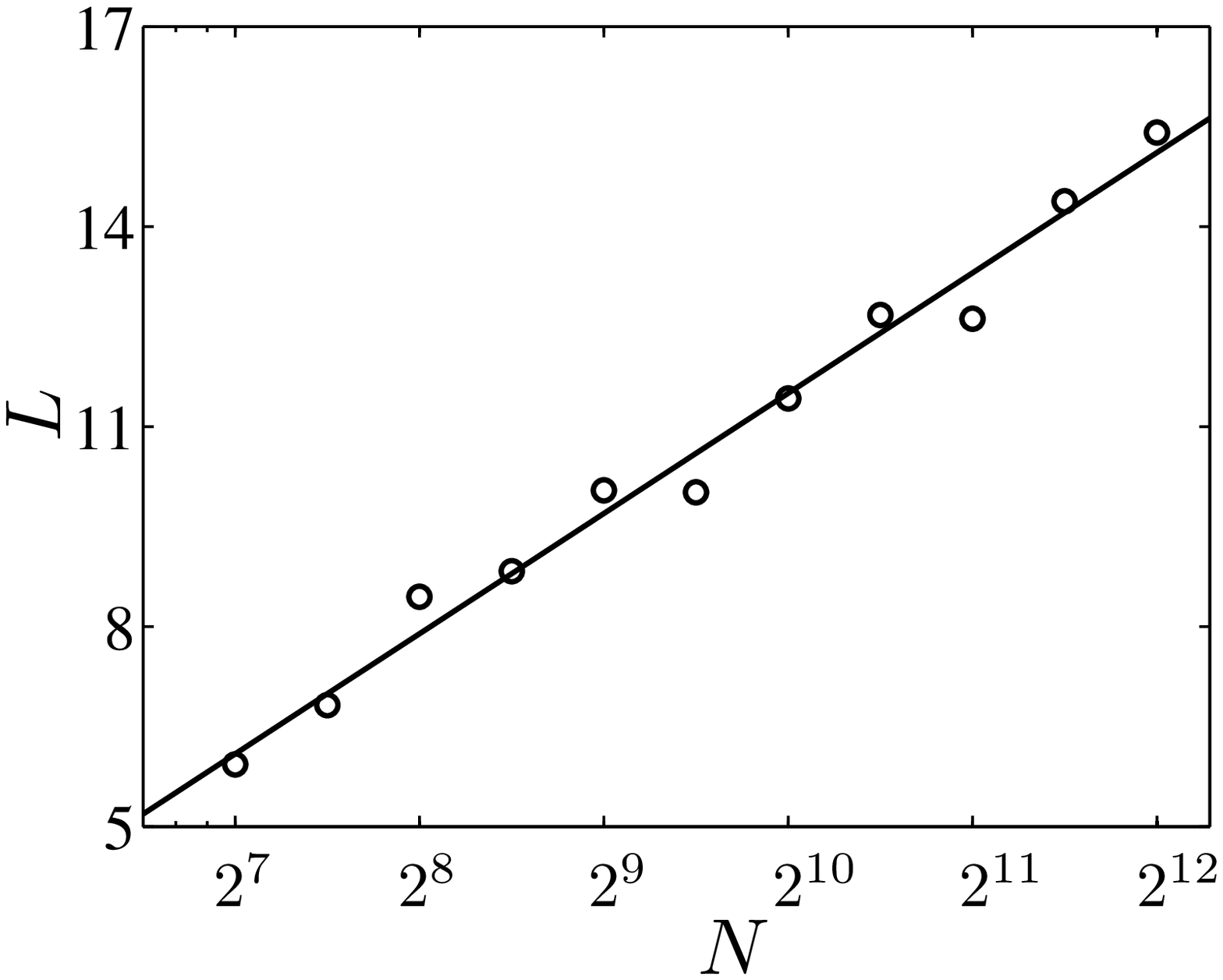}
  \includegraphics[width=8cm]{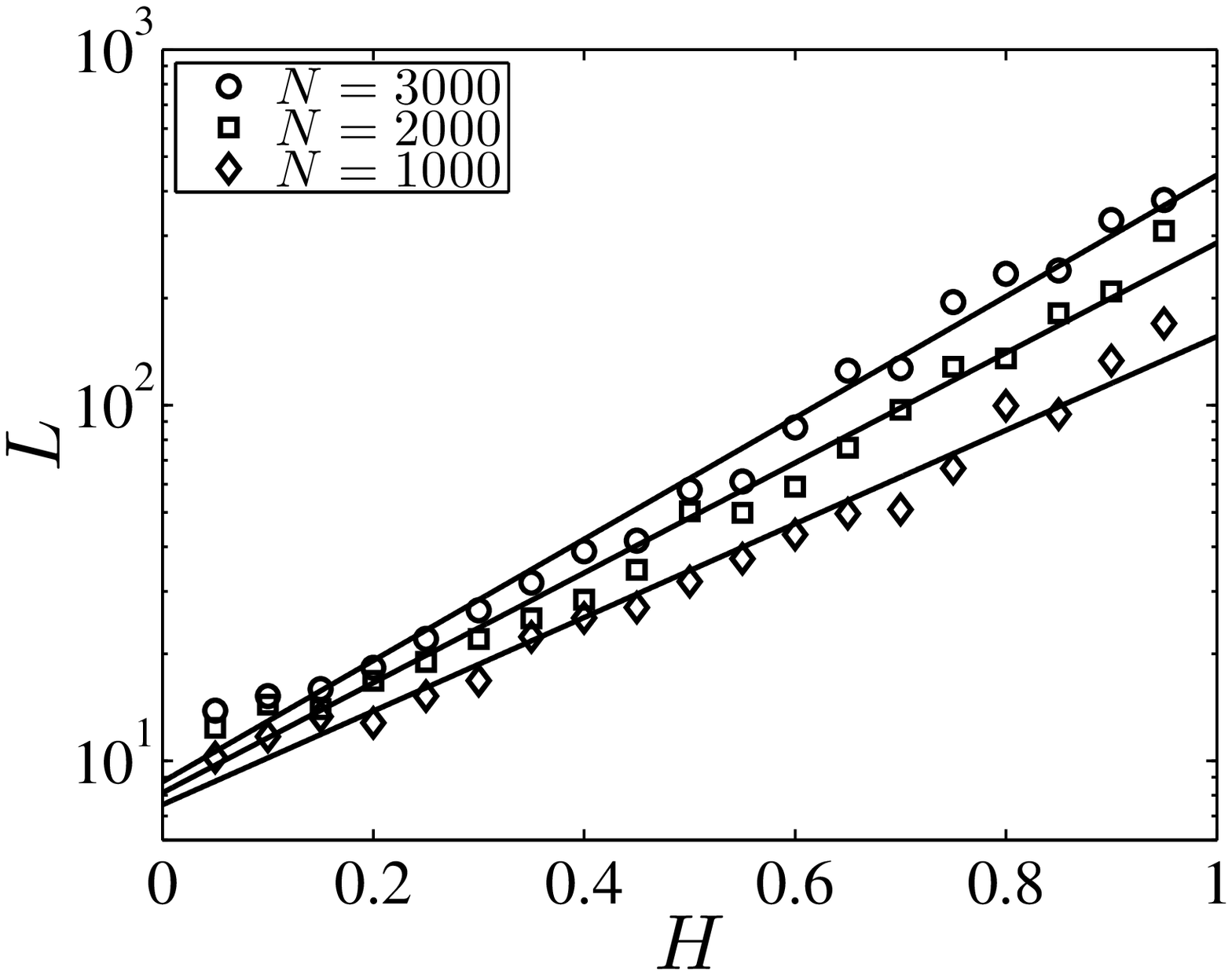}
  \caption{\label{Fig:HVG:L} (a) Logarithmic dependence of the mean length of the shortest paths ($L$) as a function of the HVG size ($N$) for FBM series with $H = 0.1$. For each $N$, 20 repeated simulations are performed and the average values of $L$ are shown. The solid line is the linear fit of $L$ versus $\ln{N}$ to the data. (b) Exponential dependence of the mean length of the shortest paths ($L$) with respect to the Hurst index $H$ for fixed HVG size. Three sizes $N=1000$, $2000$ and 3000 are illustrated. The solid lines are the linear fits of $\ln{L}$ versus $H$ to the three data sets.}
\end{figure}

According to Fig.~\ref{Fig:HVG:Examples}, when $H$ is close to 1, the HVG looks like a chain, which implies that the mean length $L$ might be proportional to the FBM length $N$. This is verified by numerical simulations. When $H$ is close to 0, we find that $L$ is proportional to $\ln{N}$, indicating that these HVGs exhibit small-world feature. Therefore, we have
\begin{equation}
  L(N) \propto
  \left\{
    \begin{array}{lll}
         N, & H\rightarrow 1, \\
     \ln N, & H\rightarrow 0.
    \end{array}
  \right.
  \label{eq:MSP2}
\end{equation}
As an example, Fig.~\ref{Fig:HVG:L}(a) illustrates the logarithmic dependence of the mean length of the shortest paths with respect to the HVG size for FBM series with $H = 0.1$.

As Fig.~\ref{Fig:HVG:Examples} suggests qualitatively, the mean length of the shortest paths increases with the Hurst index. We attempt to determine the quantitative relationship between $L$ and $H$ based on numerical simulations. The $L$ values for different $H$ are determined for fixed graph size $N$. Figure \ref{Fig:HVG:L}(b) plots $L$ with respect to $H$ in semi-logarithmic scales for $N=1000$, 2000 and 3000. It is observed that, for fixed $N$, $L$ increases exponentially in regard to $H$: 
\begin{equation}
  L(H) \sim e^{\beta H},
  \label{Eq:HVG:L:H}
\end{equation}
where $\beta$ is a function of $N$ and independent of $H$. Figure \ref{Fig:HVG:L}(b) indicates that $\beta$ increases with $N$.

\subsection{Motif distribution}
\label{S2:MotifDistribution}

Complex networks can be classified into different universality classes based on their macroscopic global properties \cite{Amaral-Scala-Barthelemy-Stanley-2000-PNAS}. Moreover, complex networks with the same global statistics may exhibit different local structure properties. At the microscopic level, the building blocks of complex networks are subgraphs or motifs and the network motif patterns of occurrence can be used to define superfamilies of networks \cite{Milo-ShenOrr-Itzkovitz-Kashtan-Chklovskii-Alon-2002-Science,Milo-Itzkovitz-Kashtan-Levitt-ShenOrr-Ayzenshtat-Sheffer-Alon-2004-Science}. The superfamily phenomenon is also observed in time series after mapping into nearest-neighbor networks, which is able
to distinguish different types of dynamics in periodic, chaotic and noisy processes based on the occurrence frequency patterns of network motifs \cite{Xu-Zhang-Small-2008-PNAS}. For nonstationary time series, the classification of superfamilies is strongly related to the DFA exponent of the time series \cite{Liu-Zhou-2010-JPA}. It is thus interesting to investigate the pattern of motif distributions in the HVGs mapped from FBMs with different Hurst indexes.

\begin{figure}[htb]
\centering
\includegraphics[width=8cm]{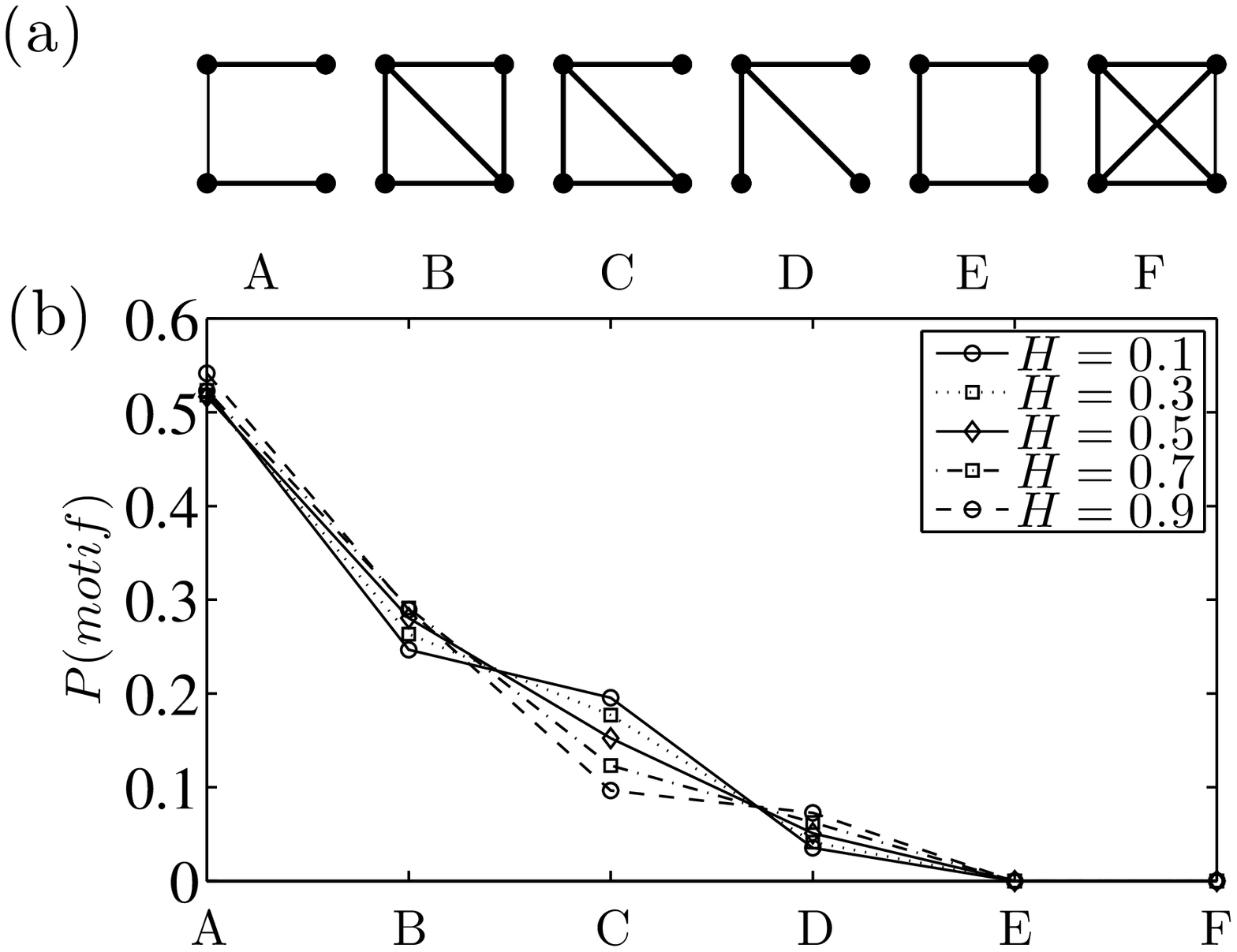}
\includegraphics[width=8cm]{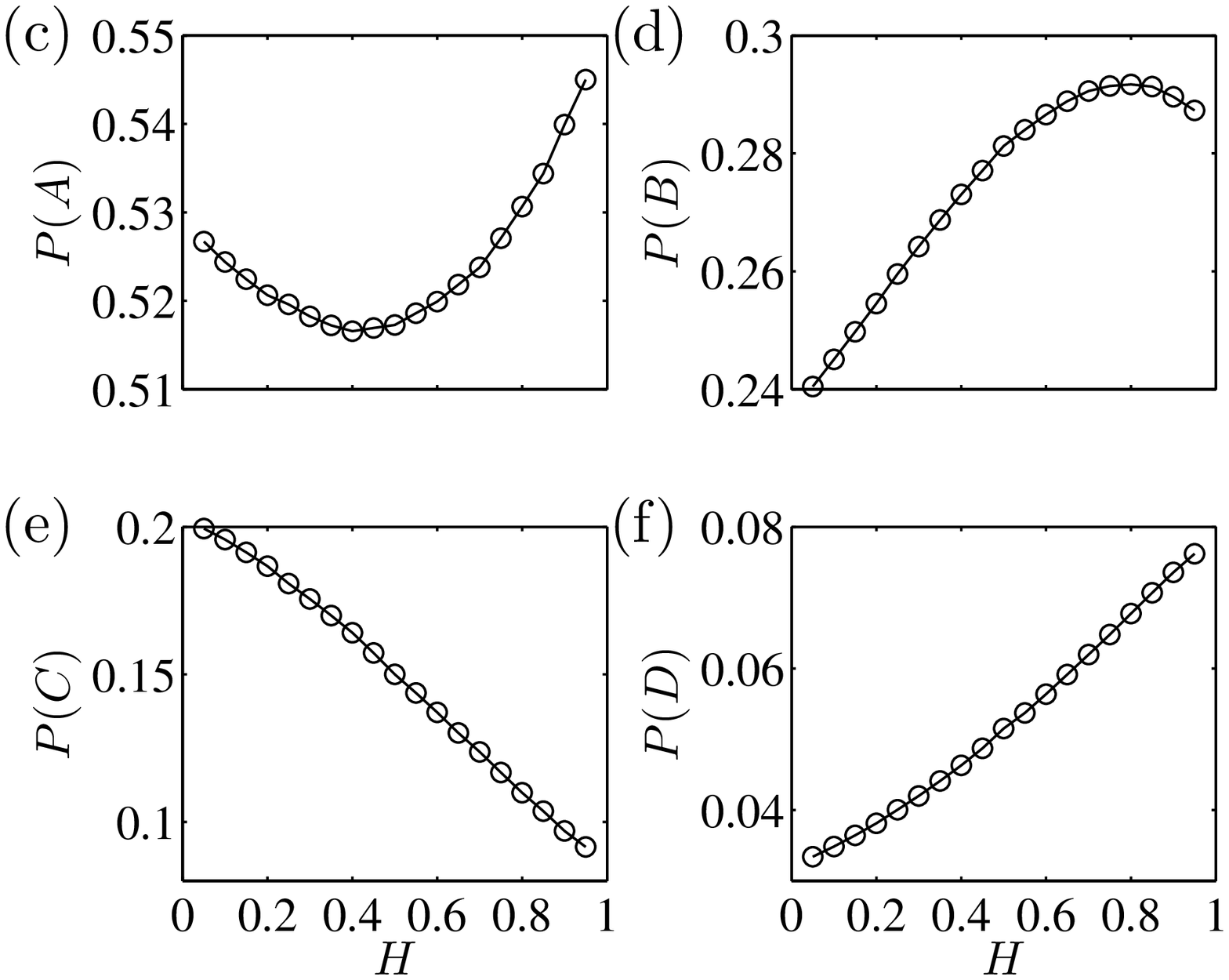}
\caption{\label{Fig:HVG:Motif} (a) All six network motifs of size 4, which are labeled $A$, $B$, $C$, $D$, $E$ and $F$, respectively. (b) Motif distributions for different FBMs with different Hurst indexes. (c) Dependence of occurrence frequency of motif $A$ as a function of $H$. (d) Same as (c) but for motif $B$. (e) Same as (c) but for motif $C$. (f) Same as (c) but for motif $D$.}
\end{figure}

Since the HVGs constructed are undirected, we consider motifs of size 4. There are six different motifs as shown in Fig.~\ref{Fig:HVG:Motif}(a). Our simulations are performed for different $H$ values. For each $H$, we generate 20 fractional Brownian motions and convert them into 20 HVGs. For each HVG, the occurrence numbers of the six motifs are determined, denoted as $n(M)$ where $M=A$, $B$, $C$, $D$, $E$ and $F$. The occurrence frequencies of the six motifs are calculated as follows
\begin{equation}
  P(M) = {n(M)}\left/{\sum_{M=A,\ldots,F} n(M)}\right..
\end{equation}
For each $H$, the average values of $P(M)$ are obtained. Figure \ref{Fig:HVG:Motif}(b) shows the motif distributions $P(M)$ for different $H$. We find that the motif distribution patterns are same for all the $H$ values:
\begin{equation}
 P(A)>P(B)>P(C)>P(D)>P(E)=P(F)=0.\nonumber
\end{equation}
For a given $H$, the motif occurrence frequencies for repeated simulations exhibit negligible fluctuations. Different from the nearest-neighbor networks \cite{Liu-Zhou-2010-JPA}, the Hurst index is not capable of distinguishing superfamilies. Plots (c-f) of Fig.~\ref{Fig:HVG:Motif} illustrate the nontrivial dependence of $P(M)$ on $H$ for $M=A$, $B$, $C$ and $D$.


The most striking feature of Fig.~\ref{Fig:HVG:Motif} is that $n(E)=n(F)=0$. In other words, the two motifs $E$ and $F$ do not occur in all the simulated HVGs. Indeed, $n(F)=0$ holds for all HVGs and $n(E)=0$ almost holds for all HVGs. We provide a proof below. Consider four arbitrary data points $\{x_{i_1}, x_{i_2},x_{i_3},x_{i_4}\}$ of a given time series, where $i_1<i_2<i_3<i_4$. For simplicity, we rewrite them as $\{x_1,x_2,x_3,x_4\}$. The four data points are mapped to four nodes of the resultant HVG.

We first explain why motif $F$ does not occur. In motif $F$, each node is connected to the other three nodes. If there are any data points between $x_1$ and $x_4$, all of them must no larger than $x_1$, $x_2$, $x_3$ and $x_4$. Otherwise, it is impossible to transform $\{x_1,x_2,x_3,x_4\}$ into a complete graph as motif $F$. Since there is a link between $x_1$ and $x_3$, the $x_2$ bar should be lower than $x_1$ and $x_3$. We have $x_2<x_3$. On the other hand, since $x_2$ and $x_4$ are visible, we have $x_3<x_2$. These two constraints contradict, which means that the two links $(x_1,x_3)$ and $(x_2,x_4)$ cannot occur simultaneously.

The explanation of $n(E) = 0$ is slightly complicated. Since each node is connected to two nodes but unconnected to one node, there are three possible cases: (1) $e_{ij}=1$ except that $e_{12}=0$ and $e_{34}=0$, (2) $e_{ij}=1$ except that $e_{13}=0$ and $e_{24}=0$, and (3) $e_{ij}=1$ except that $e_{14}=0$ and $e_{23}=0$. For cases (1) and (3), $e_{13}=1$ and $e_{24}=1$. According to Eq.~(\ref{Eq:HVG}), we have $x_3>x_2$ from $e_{13}=1$ and $x_2>x_3$ from $e_{24}=1$. Therefore, these two cases are forbidden. In case (2), we have $e_{14}=1$, $e_{13}=0$ and $e_{24}=0$. Since $e_{12}=1$, all data points between $x_1$ and $x_2$ are less than $x_1$ and $x_2$. Similarly, since $e_{23}=1$, all points between $x_2$ and $x_3$ are less than $x_2$ and $x_3$. The fact that $e_{13}=0$ results in $x_1\leq x_2$ or $x_3\leq x_2$. In addition, we have $x_1>x_2$ according to $e_{14}=1$. We thus have $x_3\leq x_2$. Similarly, when we consider $e{14}=1$ and $e_{24}=0$, we have $x_2\leq x_3$. At last, we obtain that $x_2=x_3$. Combining all the information, motif $E$ occurs if and only if $x_1,x_4>x_2=x_3$ and $e_{12}=e_{23}=e_{34}=1$. The absence of motif $E$ in Fig.~\ref{Fig:HVG:Motif}(b) simply means that the condition is not easy to be fulfilled for FBMs.

\section{Fractal scaling}
\label{S1:Fractal}

It is well known that many complex networks are small worlds \cite{Watts-Strogatz-1998-Nature}. Interestingly, small-world networks may also possess self-similar topological structure \cite{Song-Havlin-Makse-2005-Nature,Gallos-Song-Makse-2007-PA}. The fractal dimension of the network can be obtained by the conventional box-counting method based on node covering \cite{Song-Havlin-Makse-2006-NPhys,Song-Gallos-Havlin-Makse-2007-JSM,Kim-Goh-Kahng-Kim-2007-Chaos,Gao-Hu-Di-2008-PRE} or edge covering \cite{Zhou-Jiang-Sornette-2007-PA}. If a network is fractal, the minimal number $N_B$ of boxes needed to cover the network scales with respect to the box size $l_B$ as \cite{Song-Havlin-Makse-2005-Nature}
\begin{equation}
  N_B(l_B) \sim l_B^{-d_B},
  \label{Eq:N:l}
\end{equation}
where $d_B$ is the fractal dimension of the network characterizing its self-similar structure \cite{Mandelbrot-1983}. On the other hand, scale-free networks may also exhibit self-similar connectivity behaviors after coarse graining processes, in which the degree distribution is scale invariant \cite{Kim-2004-PRL,Goh-Salvi-Kahng-Kim-2006-PRL,Kim-Goh-Salvi-Oh-Kahng-Kim-2007-PRE}. Note that fractal scaling and self-similar connectivity behavior are two different properties \cite{Kim-Goh-Kahng-Kim-2007-NJP}. In this section, we investigate the fractal nature of the HVGs based on the node-covering method.

\subsection{Box-counting method based on node covering using the simulated annealing algorithm}
\label{S2:NodeCovering}

According to the definition of fractal dimensions, we need to determine the minimal number of boxes of size $l_B$ to cover the nodes of a network. We use the simulated annealing algorithm \cite{Kirkpatrick-Gelatt-Vecchi-1983-Science,Basu-Frazer-1990-Science} to determinate the minimum number $N_B$ of boxes of size $l_B$ needed to cover all the nodes of a network. The simulated annealing algorithm was used in the edge-covering method for the determination of the fractal dimension \cite{Zhou-Jiang-Sornette-2007-PA}. The simulated annealing algorithm is implemented as follows.

Starting from a given partition with $C$ boxes of sizes no larger than $l_{B}$, we consider three types of operations to transform the partition into a new one: (1) One node is moved from one box with at least two nodes in it to another box if the diameters of both new boxes do not exceed $l_{B}$; (2) One node is moved out of one box with at least two nodes to form a new box consisting of one node; and (3) Two boxes merge to form a new box, a move which is allowed if the diameter of the resulting box is no larger than $l_{B}$.

At each temperature $T$, we perform $k_{1}$ operations of the first type to exchange nodes, $k_{2}$ operations of the second type to form new boxes, and $k_{3}$ operations of the third type to merge boxes. An operation is accepted with probability $p$, where
\begin{equation}
  p=\left\{
  \begin{array}{ccccc}
    1, & C_{a}\leq C_{b} \\
    \exp[-(C_{a}-C_{b})/T], & C_{a}>C_{b}
  \end{array}
  \right.,
  \label{Eq:SimAnneal}
\end{equation}
where $C_{b}$ and $C_{a}$ are the numbers of boxes {\em{before}} and {\em{after}} an operation. It is obvious that an operation is more likely to be accepted at high temperature. After $k_1 + k_2 + k_3$ possible operations are performed, the system is cooled down to a lower temperature $T'= cT$, where $c$ is a constant less than and close to 1 (typically $c = 0.995\sim0.999$). The typical starting temperature is around $T =0.6$. If the number of boxes do not change after the $k_1 + k_2 + k_3$ operations at certain temperature, we argue that the algorithm reaches the optimal coverage and cease the annealing process.

\subsection{Fractal dimension $d_B$}
\label{S2:dB}

\begin{figure}[htb]
  \centering
  \includegraphics[width=8cm]{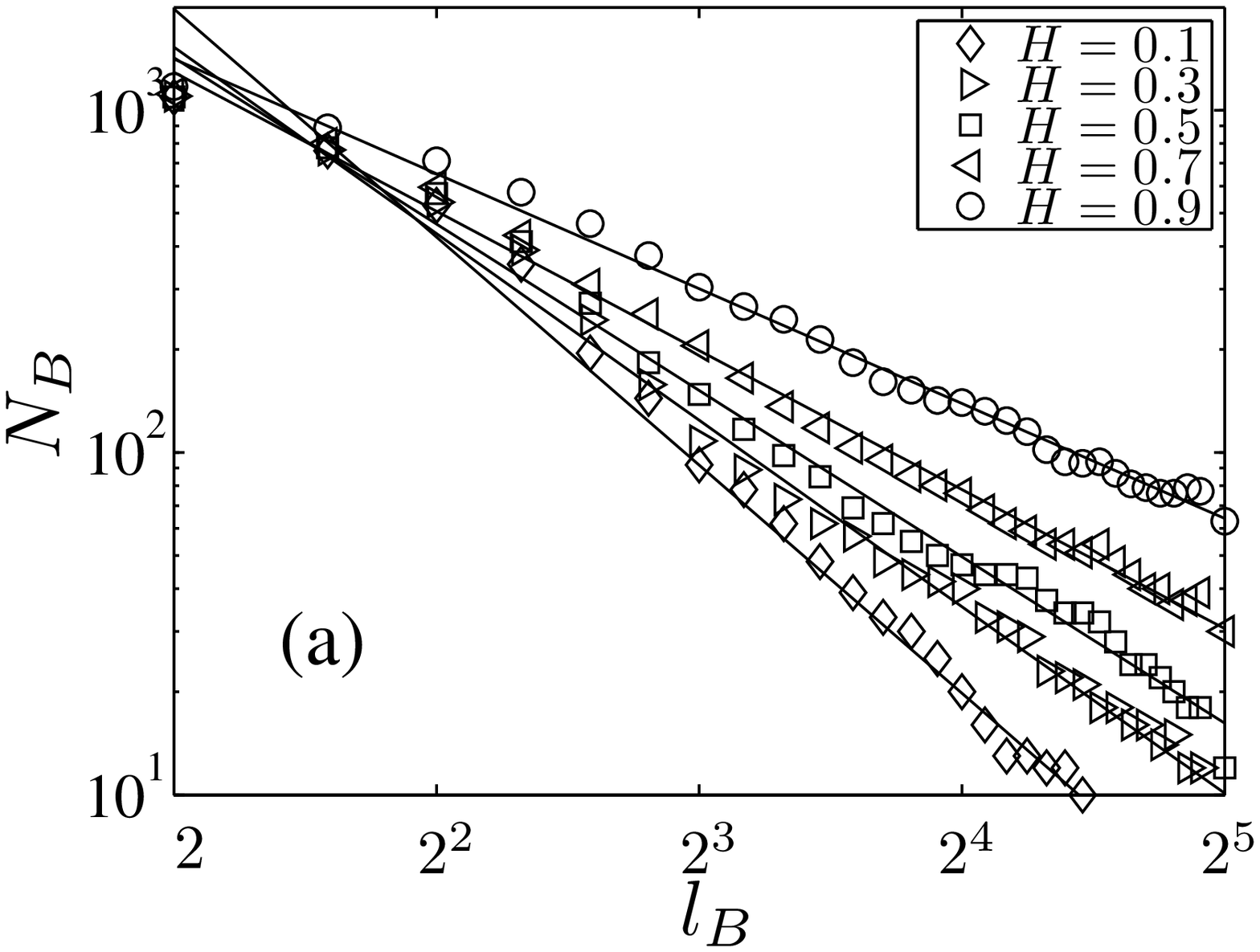}
  \includegraphics[width=8cm]{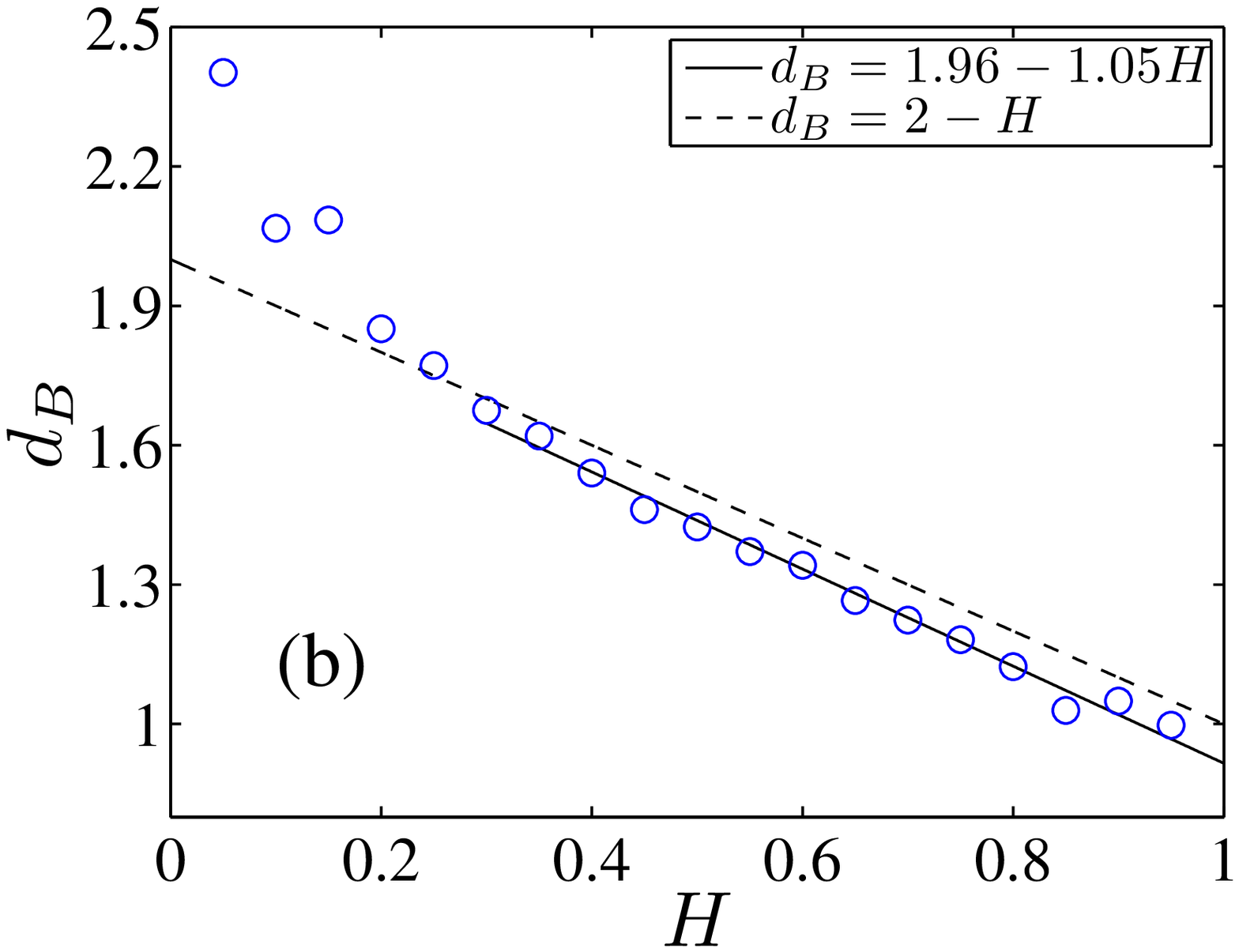}
  \caption{\label{Fig:HVG:dB} Fractal analysis of the HVGs. (a) Log-log plot of box number $N_B$ against box size $l_B$ for different FBM series with Hurst indexes $H = 0.1$, 0.3, 0.5, 0.7, and 0.9, respectively. The solid lines are best fits to the data for different Hurst indexes. (b) Linear relation between the fractal dimension $d_{B}$ and the Hurst index $H$ when $H\geq0.3$. Ten repeated calculations for each $H$ are performed and the results are the averages. The dashed line is $d_B=2-H$ for comparison.}
\end{figure}

We synthesize FBMs of size $3\times10^3$ for different Hurst indexes. Each FBM is converted into a HVG. For different box sizes $l_B$, the minimal numbers $N_B$ of boxes are determined using the node-covering method through simulated annealing as described in Section \ref{S2:NodeCovering}. It is found for all HVGs investigated that $N_B$ scales as a power law with respect to $l_B$, as expressed in Eq.~(\ref{Eq:N:l}). Figure \ref{Fig:HVG:dB}(a) illustrates four typical examples of the nice power-law dependence with $H=0.1$, 0.3, 0.5, 0.7 and 0.9. The linear regression of $\ln{N_B}$ against $\ln{l_B}$ for each Hurst index gives an estimate of $-d_B$. For each Hurst index, 10 FBMs are generated and 10 values of $d_B$ are determined. The dependence of the fractal dimension $d_B$ averaged over 10 realizations with respect to the Hurst index $H$ is depicted in Fig.~\ref{Fig:HVG:dB}(b). We observe that the fractal dimension decreases with increasing $H$, which is consistent with the HVG structures shown in Fig.~\ref{Fig:HVG:Examples}.

From Fig.~\ref{Fig:HVG:dB}(b), there is roughly a linear relation between $d_B$ and $H$ when the Hurst index is larger than 0.3, and a linear regression gives that
\begin{equation}
  d_{B} = 1.96-1.05H.
\end{equation}
The error bar of the coefficient of the linear term is 0.03 and the corresponding $p$-value is zero according to the t-test, showing that the coefficient is significantly different from zero and the linearity is confirmed. When $H$ is very close to 1, we have $d_B \approx 0.91$. It means that the HVG has a chain-like structure. When $H$ is close to 0, we have $d_B\approx 1.96$, which is however less than the extrapolated empirical value. At a first glance, it is surprising that the fractal dimension exceeds 2 while the HVGs are planar graphs. This reflects the fact that the planar HVGs for small $H$'s have to ruffle in the three- or higher-dimensional space so that the distance between two connected nodes is a finite constant while the distance between two unconnected nodes is infinite. It is thus possible to observe fractal dimensions larger than 2. For large Hurst indexes, it is interesting to compare the data with the following equation
\begin{equation}
  d_{B} \approx 2-H.
\end{equation}
It means that the fractal dimensions of these HVGs are close to, but less than, the fractal dimensions of the associated FBMs. This finding is consistent with the chain-like structures of the HVGs with large Hurst indexes.

\section{Assortative mixing pattern}
\label{S1:Assortative}

Mixing patterns are important statistical characteristics of complex networks \cite{Newman-2002-PRL,Newman-2003-PRE}. Assortative mixing refers to the phenomenon that nodes in a network are more likely connected to nodes that are alike them in some way. The mixing pattern can be measured using the Pearson coefficient
\begin{equation}
 r=\frac{M^{-1}\sum_{i}j_{i}k_{i}-\left[M^{-1}\sum_{i}\frac{1}{2}(j_{i}+k_{i})\right]^{2}}
        {M^{-1}\sum_{i}\frac{1}{2}(j_{i}^{2}+k_{i}^{2})-\left[M^{-1}\sum_{i}\frac{1}{2}(j_{i}+k_{i})\right]^{2}},
\end{equation}
where $M$ is the total number of edges, $i=1, \ldots, M$, and $j_i$ and $k_i$ denote the degrees of the two vertices at the ends of
edge~$i$. A network is assortative if $r>0$ and disassortative if $r<0$.

\begin{figure}[htb]
  \centering
  \includegraphics[width=8cm]{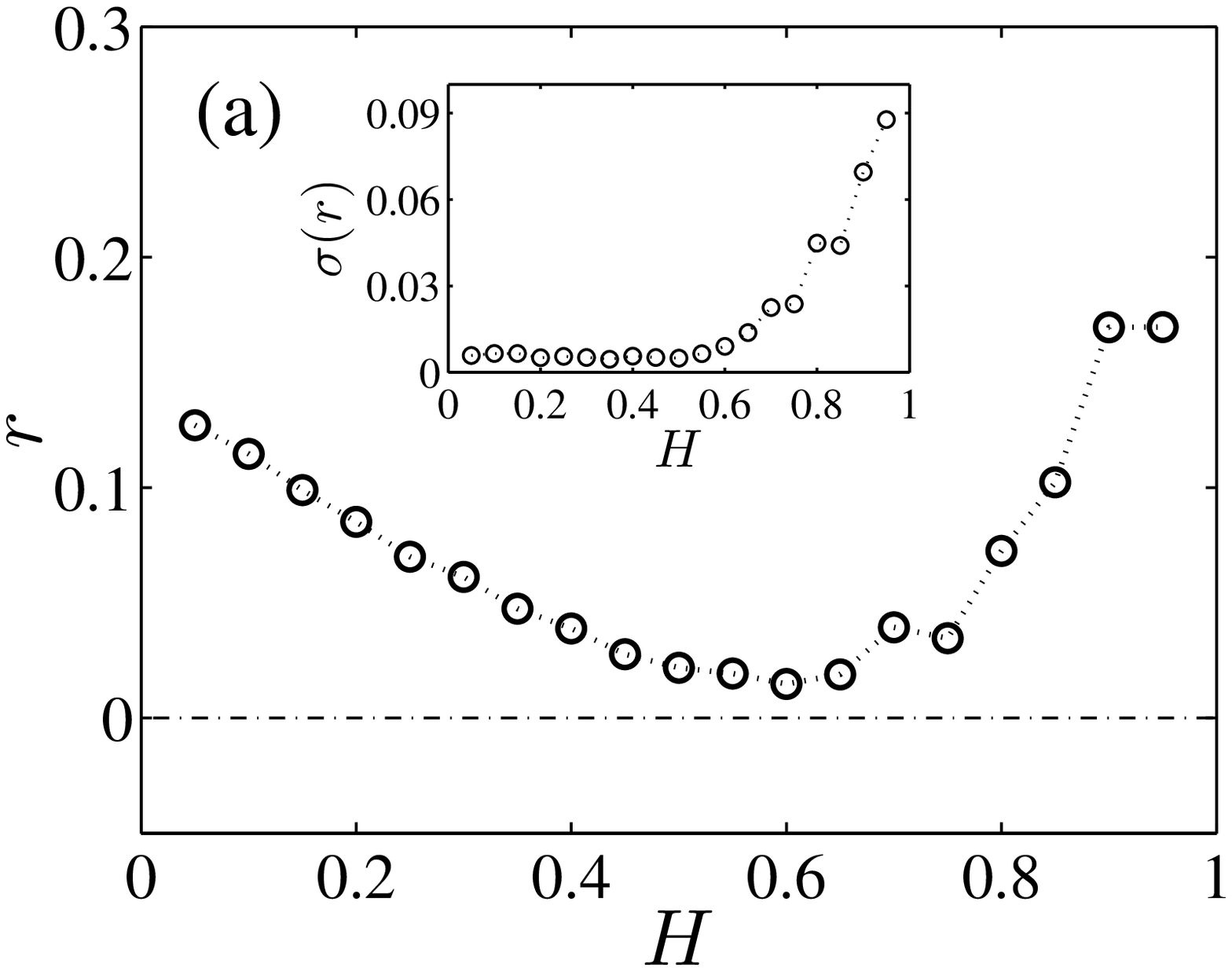}
  \includegraphics[width=8cm]{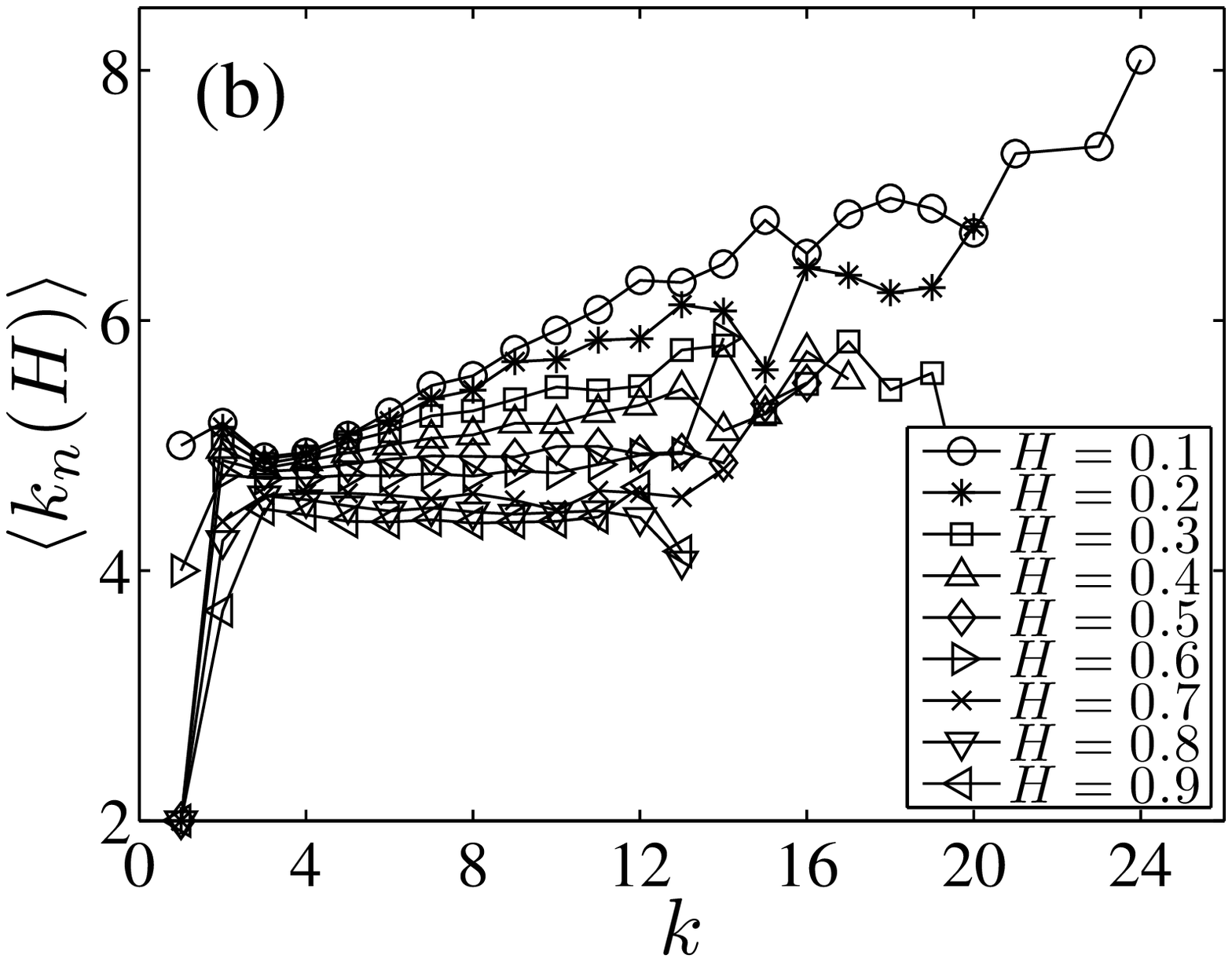}
  \caption{\label{Fig:HVG:Mixing} Mixing pattern of HVGs. (a) Pearson coefficients $r$ of HVGs mapped from different FBM series with different Hurst indexes $H$. The length of each FBM series is $10^4$ and we simulated 30 times for each Hurst index $H$. Each data point represents the average value over the 30 repeated simulations. The inset shows the standard deviation $\sigma(r)$ of the Pearson coefficient $r$ for different FBM series with different Hurst indexes $H$. (b) Dependence of the average degree $\langle k_{n}(H)\rangle$ of the neatest neighbors of nodes of the same degree $k$ with respect to $k$ for different $H$ values.}
\end{figure}

We synthesize FBMs of size $N=10^4$ with different Hurst indexes and transform them into HVGs. For each $H$, 30 FBMs are simulated and 30 Pearson coefficients are calculated. The dependence of the average Pearson coefficient on the Hurst index is illustrated in Fig.~\ref{Fig:HVG:Mixing}(a). We find that $r>0$ for all $H$ values, which is statistically significant in one standard deviation. In other words, we have $r-\sigma_r>0$ statistically. Therefore, almost all the HVGs mapped from FBMs are assortatively mixed with a few exceptions.

An alternative measure of mixing patterns is the degree correlation \cite{Newman-2003-SIAMR}. For a given network, the average degree $\langle k_{n}(H)\rangle$ of the nearest neighbors of nodes having degree $k$ is determined. If $\langle k_{n}(H)\rangle$ increases with $k$, high-degree nodes associate preferentially with other high-degree nodes and the network is assortative. Figure \ref{Fig:HVG:Mixing}(b) shows the dependence of $\langle k_{n}(H)\rangle$ with respect to $k$ for different Hurst indexes. When $H\leq0.6$, $\langle k_{n}(H)\rangle$ increases linearly with $k$ when $k\geq3$, which means that high-degree nodes are more likely to have high-degree neighbors. In addition, the slope of curve increases with decreasing $H$, which is consistent with the observation in Fig.~\ref{Fig:HVG:Mixing}(a) that $r$ decreases with $H$. For the curves with $H>0.6$, $\langle k_{n}(H)\rangle$ increases when $k\leq4$ and remains roughly constant when $k>4$, which means that low-degree nodes are more likely to have low-degree neighbors. In other words, high-degree nodes with $k>4$ are homogenously mixed, while low-degree nodes are assortatively mixed. This explains the large standard deviations of the Pearson coefficients for large $H$ shown in the inset of Fig.~\ref{Fig:HVG:Mixing}(a).

For independent and identically distributed random time series, the average value of the degree $k(x)$ of a node is a monotonically increasing function of the datum height $x$ \cite{Luque-Lacasa-Ballesteros-Luque-2009-PRE}. Similar phenomenon is observed for fractional Brownian motions with different functional form, especially for $H\leq 0.6$. Therefore, high-degree nodes are more visible to each other, leading to the assortative mixing pattern in HVGs.

\section{Summary}
\label{S1:Conclusion}

In recent years, ideas and tools in complex network sciences have been utilized to perform nonlinear time series analysis. A time series is transformed into networks based on various algorithms so that the dynamics of the process are mapped into the topological properties of the associated networks. A recently proposed algorithm based on horizontal visibility of data points is investigated in this work using fractional Brownian motions.

We have studies the topological properties of the horizontal visibility graphs constructed from FBMs with different Hurst index $H\in(0,1)$. It is found that the degree distribution, the clustering coefficient and the mean length $L$ of the shortest paths all depend on the Hurst index.
Specifically, with the increase of the Hurst index $H$, the absolute slope $\lambda$ of the exponential degree distribution increases, the clustering coefficient $C$ decreases, and the mean length $L$ of the shortest paths increases exponentially. In contrast, although the occurrence of different motifs changes with $H$, the motif rank pattern remains unchanged for different $H$.

We also performed fractal analysis using the node-covering box-counting method based on the simulated annealing algorithm. The horizontal visibility graphs are found to be fractals and the fractal dimension $d_B$ decreases with $H$. Furthermore, the HVGs are found to exhibit assortative mixing patterns because the Pearson coefficients of the networks are positive and the degree-degree correlations increase with the degree. The assortativity of HVGs means that high-degree nodes are more likely connected with other high-degree nodes, especially for $H<0.6$. This feature cannot be explained by the hub-hub repulsion mechanism of fractality in networks \cite{Song-Havlin-Makse-2006-NPhys}. Our investigations show that hub-hub attraction can also produce fractality. It seems that the HVGs mapped from FBMs have different growth mechanisms compared with many real networks in which hub-hub attraction leads to non-fractality and assortativity and hub-hub repulsion leads to fractality and disassortativity \cite{Song-Havlin-Makse-2006-NPhys}.

\bigskip
{\textbf{Acknowledgments:}}

This work was partially supported by the National Natural Science Foundation of China under grant no. 11075054 and the Fundamental Research Funds for the Central Universities.

\bibliography{E:/papers/Auxiliary/Bibliography}

\begin{thebibliography}{10}
\expandafter\ifx\csname url\endcsname\relax
  \def\url#1{\texttt{#1}}\fi
\expandafter\ifx\csname urlprefix\endcsname\relax\def\urlprefix{URL }\fi
\expandafter\ifx\csname href\endcsname\relax
  \def\href#1#2{#2} \def\path#1{#1}\fi

\bibitem{Small-Zhang-Xu-2009-LNICST}
X.-K. Xu, J.~Zhang, M.~Small, {Transforming time series into complex networks},
  Lect. Notes Institute Comput. Sci. Soc. Infor. Telecommun. Engineering 5
  (2009) 2078--2089.
\newblock \href {http://dx.doi.org/10.1007/978-3-642-02469-6\_84}
  {\path{doi:10.1007/978-3-642-02469-6\_84}}.

\bibitem{Donner-Small-Donges-Marwan-Zou-Xiang-Kurths-2011-IJBC}
R.~V. Donner, M.~Small, J.~F. Donges, N.~Marwan, Y.~Zou, R.-X. Xiang,
  J.~Kurths, {Recurrence-based time series analysis by means of complex network
  methods}, Int. J. Bifur. Chaos 21 (2011) in press.

\bibitem{Albert-Barabasi-2002-RMP}
R.~Albert, A.-L. Barab{\'a}si, {Statistical mechanics of complex networks},
  Rev. Mod. Phys. 74 (2002) 47--97.
\newblock \href {http://dx.doi.org/10.1103/RevModPhys.74.47}
  {\path{doi:10.1103/RevModPhys.74.47}}.

\bibitem{Newman-2003-SIAMR}
M.~E.~J. Newman, The structure and function of complex networks, SIAM Rev.
  45~(2) (2003) 167--256.
\newblock \href {http://dx.doi.org/10.1137/S003614450342480}
  {\path{doi:10.1137/S003614450342480}}.

\bibitem{Boccaletti-Latora-Moreno-Chavez-Hwang-2006-PR}
S.~Boccaletti, V.~Latora, Y.~Moreno, M.~Chavez, D.-U. Hwang, {Complex networks:
  Structure and dynamics}, Phys. Rep. 424 (2006) 175--308.
\newblock \href {http://dx.doi.org/10.1016/j.physrep.2005.10.009}
  {\path{doi:10.1016/j.physrep.2005.10.009}}.

\bibitem{Zhang-Small-2006-PRL}
J.~Zhang, M.~Small, {Complex network from pseudoperiodic time series: Topology
  versus dynamics}, Phys. Rev. Lett. 96 (2006) 238701.
\newblock \href {http://dx.doi.org/10.1103/PhysRevLett.96.238701}
  {\path{doi:10.1103/PhysRevLett.96.238701}}.

\bibitem{Zhang-Sun-Luo-Zhang-Nakamura-Small-2008-PD}
J.~Zhang, J.-F. Sun, X.-D. Luo, K.~Zhang, T.~Nakamura, M.~Small,
  {Characterizing pseudoperiodic time series through the complex network
  approach}, Physica D 237 (2008) 2856--2865.

\bibitem{Zhang-Zhou-Wang-2010-PProc}
J.-H. Zhang, H.-X. Zhou, Y.-G. Wang, {Network topologies of Shanghai stock
  index}, Phys. Proc. 3 (2010) 1733--1740.
\newblock \href {http://dx.doi.org/10.1016/j.phpro.2010.07.012}
  {\path{doi:10.1016/j.phpro.2010.07.012}}.

\bibitem{Li-Wang-2006-CSB}
P.~Li, B.-H. Wang, {An approach to Hang Seng Index in Hong Kong stock market
  based on network topological statistics}, Chinese Science Bulletin 51 (2006)
  624--629.
\newblock \href {http://dx.doi.org/10.1007/s11434-006-0624-4}
  {\path{doi:10.1007/s11434-006-0624-4}}.

\bibitem{Li-Wang-2007-PA}
P.~Li, B.-H. Wang, {Extracting hidden fluctuation patterns of Hang Seng stock
  index from network topologies}, Physica A 378 (2007) 519--526.
\newblock \href {http://dx.doi.org/10.1016/j.physa.2006.10.089}
  {\path{doi:10.1016/j.physa.2006.10.089}}.

\bibitem{Li-Yang-Komatsuzak-2008-PNAS}
C.-B. Li, H.~Yang, T.~Komatsuzaki, {Multiscale complex network of protein
  conformational fluctuations in single-molecule time series}, Proc. Natl.
  Acad. Sci. U.S.A. 105 (2008) 536--541.
\newblock \href {http://dx.doi.org/10.1073/pnas.0707378105}
  {\path{doi:10.1073/pnas.0707378105}}.

\bibitem{Lacasa-Luque-Ballesteros-Luque-Nuno-2008-PNAS}
L.~Lacasa, B.~Luque, F.~Ballesteros, J.~Luque, J.~C. Nu{\~n}o, {From time
  series to complex networks: The visibility graph}, Proc. Natl. Acad. Sci.
  U.S.A. 105 (2008) 4972--4975.
\newblock \href {http://dx.doi.org/10.1073/pnas.0709247105}
  {\path{doi:10.1073/pnas.0709247105}}.

\bibitem{Luque-Lacasa-Ballesteros-Luque-2009-PRE}
B.~Luque, L.~Lacasa, F.~Ballesteros, J.~Luque, {Horizontal visibility graphs:
  Exact results for random time series}, Phys. Rev. E 80 (2009) 046103.
\newblock \href {http://dx.doi.org/10.1103/PhysRevE.80.046103}
  {\path{doi:10.1103/PhysRevE.80.046103}}.

\bibitem{Xu-Zhang-Small-2008-PNAS}
X.-K. Xu, J.~Zhang, M.~Small, {Superfamily phenomena and motifs of networks
  induced from time series}, Proc. Natl. Acad. Sci. U.S.A. 105 (2008)
  19601--19605.
\newblock \href {http://dx.doi.org/10.1073/pnas.0806082105}
  {\path{doi:10.1073/pnas.0806082105}}.

\bibitem{Gao-Jin-2009-Chaos}
Z.-K. Gao, N.-D. Jin, {Complex network from time series based on phase space
  reconstruction}, Chaos 19 (2009) 033137.
\newblock \href {http://dx.doi.org/10.1063/1.3227736}
  {\path{doi:10.1063/1.3227736}}.

\bibitem{Liu-Zhou-2010-JPA}
C.~Liu, W.-X. Zhou, Superfamily classification of nonstationary time series
  based on dfa scaling exponents, J. Phys. A: Math. Theor. 43 (2010) 495005.
\newblock \href {http://dx.doi.org/10.1088/1751-8113/43/49/495005}
  {\path{doi:10.1088/1751-8113/43/49/495005}}.

\bibitem{Yang-Yang-2008-PA}
Y.~Yang, H.-J. Yang, {Complex network-based time series analysis}, Physica A
  387 (2008) 1381--1386.
\newblock \href {http://dx.doi.org/10.1016/j.physa.2007.10.055}
  {\path{doi:10.1016/j.physa.2007.10.055}}.

\bibitem{Gao-Jin-2009-PRE}
Z.-K. Gao, N.-D. Jin, {Flow-pattern identification and nonlinear dynamics of
  gas-liquid two-phase flow in complex networks}, Phys. Rev. E 79 (2009)
  066303.
\newblock \href {http://dx.doi.org/10.1103/PhysRevE.79.066303}
  {\path{doi:10.1103/PhysRevE.79.066303}}.

\bibitem{Shirazi-Jafari-Davoudi-Peinke-Tabar-Sahimi-2009-JSM}
A.~H. Shirazi, G.~R. Jafari, J.~Davoudi, J.~Peinke, M.~R.~R. Tabar, M.~Sahimi,
  {Mapping stochastic processes onto complex networks}, J. Stat. Mech. (2009)
  P07046\href {http://dx.doi.org/10.1088/1742-5468/2009/07/P07046}
  {\path{doi:10.1088/1742-5468/2009/07/P07046}}.

\bibitem{Kostakos-2009-PA}
V.~Kostakos, {Temporal graphs}, Physica A 388 (2009) 1007--1023.
\newblock \href {http://dx.doi.org/10.1016/j.physa.2008.11.021}
  {\path{doi:10.1016/j.physa.2008.11.021}}.

\bibitem{Marwan-Donges-Zou-Donner-Kurths-2009-PLA}
N.~Marwan, J.~F. Donges, Y.~Zou, R.~V. Donner, J.~Kurths, {Complex network
  approach for recurrence analysis of time series}, Phys. Lett. A 373 (2009)
  4246--4254.
\newblock \href {http://dx.doi.org/10.1016/j.physleta.2009.09.042}
  {\path{doi:10.1016/j.physleta.2009.09.042}}.

\bibitem{Donner-Zou-Donges-Marwan-Kurths-2010-PRE}
R.~V. Donner, Y.~Zou, J.~F. Donges, N.~Marwan, J.~Kurths, {Ambiguities in
  recurrence-based complex network representations of time series}, Phys. Rev.
  E 81 (2010) 015101(R).
\newblock \href {http://dx.doi.org/10.1103/PhysRevE.81.015101}
  {\path{doi:10.1103/PhysRevE.81.015101}}.

\bibitem{Donner-Zou-Donges-Marwan-Kurths-2010-NJP}
R.~V. Donner, Y.~Zou, J.~F. Donges, N.~Marwan, J.~Kurths, {Recurrence networks:
  A novel paradigm for nonlinear time series analysis}, New J. Phys. 12 (2010)
  033025.
\newblock \href {http://dx.doi.org/10.1088/1367-2630/12/3/033025}
  {\path{doi:10.1088/1367-2630/12/3/033025}}.

\bibitem{Sinatra-Condorelli-Latora-2010-PRL}
R.~Sinatra, D.~Condorelli, V.~Latora, {Networks of motifs from sequences of
  symbols}, Phys. Rev. Lett. 105 (2010) 178702.
\newblock \href {http://dx.doi.org/10.1103/PhysRevLett.105.178702}
  {\path{doi:10.1103/PhysRevLett.105.178702}}.

\bibitem{Ni-Jiang-Zhou-2009-PLA}
X.-H. Ni, Z.-Q. Jiang, W.-X. Zhou, {Degree distributions of the visibility
  graphs mapped from fractional Brownian motions and multifractal random
  walks}, Phys. Lett. A 373 (2009) 3822--3826.
\newblock \href {http://dx.doi.org/10.1016/j.physleta.2009.08.041}
  {\path{doi:10.1016/j.physleta.2009.08.041}}.

\bibitem{Qian-Jiang-Zhou-2010-JPA}
M.-C. Qian, Z.-Q. Jiang, W.-X. Zhou, Universal and nonuniversal allometric
  scaling behaviors in the visibility graphs of world stock market indices, J.
  Phys. A: Math. Theor. 43 (2010) 335002.
\newblock \href {http://dx.doi.org/10.1088/1751-8113/43/33/335002}
  {\path{doi:10.1088/1751-8113/43/33/335002}}.

\bibitem{Lacasa-Luque-Luque-Nuno-2009-EPL}
L.~Lacasa, B.~Luque, J.~Luque, J.~C. Nu{\~n}o, {The visibility graph: A new
  method for estimating the Hurst exponent of fractional Brownian motion}, EPL
  (Europhys. Lett.) 86 (2009) 30001.
\newblock \href {http://dx.doi.org/10.1209/0295-5075/86/30001}
  {\path{doi:10.1209/0295-5075/86/30001}}.

\bibitem{Elsner-Jagger-Fogarty-2009-GRL}
J.~B. Elsner, T.~H. Jagger, E.~A. Fogarty, {Visibility network of United States
  hurricanes}, Geophys. Res. Lett. 36 (2009) L16702.
\newblock \href {http://dx.doi.org/10.1029/2009GL039129}
  {\path{doi:10.1029/2009GL039129}}.

\bibitem{Yang-Wang-Yang-Mang-2009-PA}
Y.~Yang, J.-B. Wang, H.-J. Yang, J.-S. Mang, {Visibility graph approach to
  exchange rate series}, Physica A 388 (2009) 4431--4437.
\newblock \href {http://dx.doi.org/10.1016/j.physa.2009.07.016}
  {\path{doi:10.1016/j.physa.2009.07.016}}.

\bibitem{Liu-Zhou-Yuan-2010-PA}
C.~Liu, W.-X. Zhou, W.-K. Yuan, Statistical properties of visibility graph of
  energy dissipation rates in three-dimensional fully developed turbulence,
  Physica A 389 (2010) 2675--2681.
\newblock \href {http://dx.doi.org/10.1016/j.physa.2010.02.043}
  {\path{doi:10.1016/j.physa.2010.02.043}}.

\bibitem{Shao-2010-APL}
Z.-G. Shao, {Network analysis of human heartbeat dynamics}, Appl. Phys. Lett.
  96 (2010) 073703.
\newblock \href {http://dx.doi.org/10.1063/1.3308505}
  {\path{doi:10.1063/1.3308505}}.

\bibitem{Dong-Li-2010-APL}
Z.~Dong, X.~Li, {Comment on ``Network analysis of human heartbeat dynamics''},
  Appl. Phys. Lett. 96 (2010) 266101.
\newblock \href {http://dx.doi.org/10.1063/1.3458811}
  {\path{doi:10.1063/1.3458811}}.

\bibitem{Ahmadlou-Adeli-Adeli-2010-JNT}
M.~Ahmadlou, H.~Adeli, A.~Adeli, {New diagnostic EEG markers of the Alzheimer's
  disease using visibility graph}, J. Neural Transm. 117 (2010) 1099--1109.
\newblock \href {http://dx.doi.org/10.1007/s00702-010-0450-3}
  {\path{doi:10.1007/s00702-010-0450-3}}.

\bibitem{Ahadpour-Sadra-2010-XXX}
S.~Ahadpour, Y.~Sadra, {Randomness criteria in binary visibility graph
  perspective}, arXiv: 1004.2189 (2010).

\bibitem{Fan-Guo-Zha-2010-XXX}
C.~Fan, J.-L. Guo, Y.-L. Zha, {Fractal analysis on human behaviors dynamics},
  arXiv: 1012.4088 (2010).

\bibitem{Tang-Liu-Liu-2010-MPLB}
Q.~Tang, J.~Liu, H.-L. Liu, {Comparison of different daily streamflow series in
  US and China, under a viewpoint of complex networks}, Mod. Phys. Lett. B 24
  (2010) 1541--1547.
\newblock \href {http://dx.doi.org/10.1142/S0217984910023335}
  {\path{doi:10.1142/S0217984910023335}}.

\bibitem{Gutin-Mansour-Severini-2010-XXX}
G.~Gutin, T.~Mansour, S.~Severini, {A characterization of horizontal visibility
  graphs and combinatorics on words}, arXiv: 1010.1850 (2010).

\bibitem{Lacasa-Toral-2010-PRE}
L.~Lacasa, R.~Toral, Description of stochastic and chaotic series using
  visibility graphs, Phys. Rev. E 82 (2010) 036120.
\newblock \href {http://dx.doi.org/10.1103/PhysRevE.82.036120}
  {\path{doi:10.1103/PhysRevE.82.036120}}.

\bibitem{Yook-Radicchi-MeyerOrtmanns-2005-PRE}
S.-H. Yook, F.~Radicchi, H.~Meyer-Ortmanns, {Self-similar scale-free networks
  and disassortativity}, Phys. Rev. E 72 (2005) 045105.
\newblock \href {http://dx.doi.org/10.1103/PhysRevE.72.045105}
  {\path{doi:10.1103/PhysRevE.72.045105}}.

\bibitem{Zhang-Zhou-Zou-2007-EPJB}
Z.-Z. Zhang, S.-G. Zhou, T.~Zou, {Self-similarity, small-world, scale-free
  scaling, disassortativity, and robustness in hierarchical lattices}, Eur.
  Phys. J. B 56 (2007) 259--271.
\newblock \href {http://dx.doi.org/10.1140/epjb/e2007-00107-6}
  {\path{doi:10.1140/epjb/e2007-00107-6}}.

\bibitem{Song-Havlin-Makse-2005-Nature}
C.-M. Song, S.~Havlin, H.~A. Makse, {Self-similarity of complex networks},
  Nature 433 (2005) 392--395.
\newblock \href {http://dx.doi.org/10.1038/nature03248}
  {\path{doi:10.1038/nature03248}}.

\bibitem{Amaral-Scala-Barthelemy-Stanley-2000-PNAS}
L.~A.~N. Amaral, A.~Scala, M.~Barthelemy, H.~E. Stanley, {Classes of
  small-world networks}, Proc. Natl. Acad. Sci. U.S.A. 97 (2000) 11149--11152.
\newblock \href {http://dx.doi.org/10.1073/pnas.200327197}
  {\path{doi:10.1073/pnas.200327197}}.

\bibitem{Milo-ShenOrr-Itzkovitz-Kashtan-Chklovskii-Alon-2002-Science}
R.~Milo, S.~Shen-Orr, S.~Itzkovitz, N.~Kashtan, D.~Chklovskii, U.~Alon,
  {Network motifs: Simple building blocks of complex networks}, Science 298
  (2002) 824--827.
\newblock \href {http://dx.doi.org/10.1126/science.298.5594.824}
  {\path{doi:10.1126/science.298.5594.824}}.

\bibitem{Milo-Itzkovitz-Kashtan-Levitt-ShenOrr-Ayzenshtat-Sheffer-Alon-2004-Science}
R.~Milo, S.~Itzkovitz, N.~Kashtan, R.~Levitt, S.~Shen-Orr, I.~Ayzenshtat,
  M.~Sheffer, U.~Alon, {Superfamilies of evolved and designed networks},
  Science 303 (2004) 1538--1542.
\newblock \href {http://dx.doi.org/10.1126/science.1089167}
  {\path{doi:10.1126/science.1089167}}.

\bibitem{Watts-Strogatz-1998-Nature}
D.~J. Watts, S.~H. Strogatz, Collective dynamics in `small-world' networks,
  Nature 393 (1998) 440--442.

\bibitem{Gallos-Song-Makse-2007-PA}
L.~K. Gallos, C.-M. Song, H.~A. Makse, {A review of fractality and
  self-similarity in complex networks}, Physica A 386 (2007) 686--691.
\newblock \href {http://dx.doi.org/10.1016/j.physa.2007.07.069}
  {\path{doi:10.1016/j.physa.2007.07.069}}.

\bibitem{Song-Havlin-Makse-2006-NPhys}
C.-M. Song, S.~Havlin, H.~A. Makse, {Origins of fractality in the growth of
  complex networks}, Nat. Phys. 2 (2006) 275--281.

\bibitem{Song-Gallos-Havlin-Makse-2007-JSM}
C.-M. Song, L.~K. Gallos, S.~Havlin, H.~A. Makse, {How to calculate the fractal
  dimension of a complex network: The box covering algorithm}, J. Stat. Mech.
  (2007) P03006\href {http://dx.doi.org/10.1088/1742-5468/2007/03/P03006}
  {\path{doi:10.1088/1742-5468/2007/03/P03006}}.

\bibitem{Kim-Goh-Kahng-Kim-2007-Chaos}
J.~S. Kim, K.-I. Goh, B.~Kahng, D.~Kim, {A box-covering algorithm for fractal
  scaling in scale-free networks}, Chaos 17 (2007) 026116.
\newblock \href {http://dx.doi.org/10.1063/1.2737827}
  {\path{doi:10.1063/1.2737827}}.

\bibitem{Gao-Hu-Di-2008-PRE}
L.~Gao, Y.-Q. Hu, Z.-R. Di, {Accuracy of the ball-covering approach for fractal
  dimensions of complex networks and a rank-driven algorithm}, Phys. Rev. E 78
  (2008) 046109.
\newblock \href {http://dx.doi.org/10.1103/PhysRevE.78.046109}
  {\path{doi:10.1103/PhysRevE.78.046109}}.

\bibitem{Zhou-Jiang-Sornette-2007-PA}
W.-X. Zhou, Z.-Q. Jiang, D.~Sornette, {Exploring self-similarity of complex
  cellular networks: The edge-covering method with simulated annealing and
  log-periodic sampling}, Physica A 375 (2007) 741--752.

\bibitem{Mandelbrot-1983}
B.~B. Mandelbrot, {The Fractal Geometry of Nature}, W. H. Freeman, New York,
  1983.

\bibitem{Kim-2004-PRL}
B.~J. Kim, {Geographical coarse graining of complex networks}, Phys. Rev. Lett.
  93 (2004) 168701.

\bibitem{Goh-Salvi-Kahng-Kim-2006-PRL}
K.-I. Goh, G.~Salvi, B.~Kahng, D.~Kim, {Skeleton and fractal scaling in complex
  networks}, Phys. Rev. Lett. 96 (2006) 018701.

\bibitem{Kim-Goh-Salvi-Oh-Kahng-Kim-2007-PRE}
J.~S. Kim, K.-I. Goh, G.~Salvi, E.~Oh, B.~Kahng, D.~Kim, {Fractality in complex
  networks: Critical and supercritical skeletons}, Phys. Rev. E 75 (2007)
  016110.
\newblock \href {http://dx.doi.org/10.1103/PhysRevE.75.016110}
  {\path{doi:10.1103/PhysRevE.75.016110}}.

\bibitem{Kim-Goh-Kahng-Kim-2007-NJP}
J.~S. Kim, K.-I. Goh, B.~Kahng, D.~Kim, {Fractality and self-similarity in
  scale-free networks}, New J. Phys. 9 (2007) 177.
\newblock \href {http://dx.doi.org/10.1088/1367-2630/9/6/177}
  {\path{doi:10.1088/1367-2630/9/6/177}}.

\bibitem{Kirkpatrick-Gelatt-Vecchi-1983-Science}
S.~Kirkpatrick, C.~D.~J. Gelatt, M.~P. Vecchi, {Optimization by simulated
  annealing}, Science 220 (1983) 671--680.

\bibitem{Basu-Frazer-1990-Science}
A.~Basu, L.~N. Frazer, {Rapid determination of the critical temperature in
  simulated annealing inversion}, Science 249 (1990) 1409--1412.

\bibitem{Newman-2002-PRL}
M.~E.~J. Newman, {Assortative mixing in networks}, Phys. Rev. Lett. 89 (2002)
  208701.
\newblock \href {http://dx.doi.org/10.1103/PhysRevLett.89.208701}
  {\path{doi:10.1103/PhysRevLett.89.208701}}.

\bibitem{Newman-2003-PRE}
M.~E.~J. Newman, Mixing patterns in networks, Phys. Rev. E 67 (2003) 026126.

\end{thebibliography}

\end{document}